\documentclass[10pt-default,11pt]{article}
\usepackage{amsmath}
\usepackage{amsfonts}
\usepackage{amssymb}
\usepackage{geometry}

\input{tcilatex}
\begin{document}

\bigskip

\bigskip {\huge \ }

{\huge Optical knots and contact geometry I. }

{\huge From Arnol'd inequality to Ranada's dyons}

\bigskip

\textbf{Arkady L. Kholodenko}

\bigskip

{\small 375 H.L.Hunter Laboratories, Clemson University, Clemson, SC
29634-0973,USA}

\bigskip

\textbf{Abstract\bigskip }

Recently there had been a great deal of activity associated with various
schemes

of \ designing both analytical and experimental methods \ describing knotted
structures

in electrodynamics and in hydrodynamics. The majority of \ works in
electrodynamics

were \ inspired by the influential paper by Ranada (1989) and its subsequent
refinements.

In this work and in its companion we reanalyze Ranada's results \ using
methods of contact

geometry and topology. Not only our analysis allows us to reproduce his major

results but, in addition, it provides opportunities for considerably
extending \ the catalog 

of known knot types. Furthermore, it allows\ to reinterpret \ both the
electric and 

magnetic charges purely topologically thus opening the possibility of
treatment of masses

and charges in Yang-Mills and gravity theories also topologically. According
to (now

proven) Thurston's geometrization conjecture complements of \ all
knots/links in $S^{3}$ 

are \ spaces \ of positive, zero or negative curvature. This means \ that
spaces around 

our topological masses/charges are curved. This fact is essential for design
of purely

topological theories \ of gravity, electromagnetism and \ strong/weak
interactions.

\bigskip

PACS numbers : 11.15 Yc; 11.27.+d; 11.30.-j; 42, 45.20 Jj; 47

\bigskip

1. \textbf{Introduction\bigskip }

\textit{1.1. Motivation and background}

\bigskip

In 1858 Herman von Helmholtz \ published a paper $($Helmholtz 1858) in which
he demonstrated that ideal incompressible constant density fluid should
contain \ vortex excitations which are stable\footnote{%
In fact, they were believed to be stable/permanent and indivisible (Hemholtz
1858, Thompson 1868)}. This result had attracted attention of Maxwell, Tait
and Thomson (Lord Kelvin). In particular, Thomson \ developed an atomic
vortex theory in which the rigidity of Hemholtz vortices was related to
supposed indivisibility of atoms. Different atoms are expected to have
different vortex types. To classify different vortex types Tait \ developed
what is now known as knot theory while Maxwell developed some applications
of \ these results by Thomson and Tait to chemical reactions. Subsequently,
in 1883, J.J.Thomson (the man who discovered the electron) wrote a monograph
entitled \ "A Treatise on the Motion of Vortex Rings" $($Thompson 1883) in
which he further developed the theory by extending it to the linked vortices
and streamlined Maxwell's results. Subsequent studies done century later
demonstrated the possibility of \ existence of \ torus knots and links in
ideal fluids (Ricca 2001). Although elementary vortices in fluids were
experimentally known for quite some time, the existence of knotted
structures, although allowed theoretically, escaped experimental
verification till 2013 (Kleckner and Irvine 2013).\ To prove the existence
of \ Helmholtz vortex structures Tait designed an apparatus capable of
producing smoke rings. The design of apparatus by Tait was so successful
that he was able to demonstrate that (Lomonaco 1996):

1.The vortex rings \ can \ exist for a long time.

2.On collision the rings were capable of scattering from each other as if
they

\ \ \ were made of rubber.

3.The rings exhibited some vibration modes around their circular form.

4.On each attempt to cut the smoke rings with a knife, the smoke rings

\ \ would wriggle around the knife without breaking.

Such a situation with vortices remained basically unchanged till 2013. It
should be noted that in 2013 the results were obtained in the lab (not
detected yet in Nature) while the theory allows\ the existence of
knotted/linked structures in Nature. \ Based on the results just presented,
it is clear that existence of knotted/linked structures depends upon:

a) the way these structures are prepared;

b) the way these structures are detected;

c) the stability of these structures from the moment they were created to

\ \ \ the moment they are detected.

Evidently, c) is mainly theoretical issue while a) and b) are experimental.
\ In this paper by extending some latest results from hydrodynamics of
incompressible ideal fluids to \ electrodynamics, we describe new classes of
\ knotted/linked structures not present in the works by Ranada (and
associates) and \ others e.g. (Kedia et al 2013). Although these new classes
of knots/links have their origins \ in contact geometry/topology (Kholodenko
2013), they should not be confused with the optical knots discussed by
Arnol'd (1986). Surely, mathematically they can be brought into
correspondence with each other.

\ The development of correspondences between different subdisciplines of
physics took place slowly and at different times. Perhaps the oldest, is the
correspondence between\ the dynamics of ideal fluids and mechanics. A bit
later the correspondence between geometrical optics and mechanics was
established (Arnol'd 1989)\footnote{%
Incidentally, Arnol'd optical knots have their origin in geometrical optics.}%
. Mechanics -hydrodynamics correspondence was exploited in great detail in
the book by Arnol'd (Arnol'd and Khesin 1998). Surprisingly, it develops the
formalism without much use of methods of contact geometry. \ This is
surprising since Arnol'd \ was the major proponent of contact geometry. In
yet another book by Arnol'd (Arnol'd 1984), in Chr.14, he stated that
"Contact geometry is playing in optics and theory of wave propagation the
same role as symplectic geometry for mechanics". \ This remark by Arnol'd
(apparently) was left unnoticed in physics \ literature. \ In this paper and
its companion we are making \ an attempt at\ eliminating the existing
deficiency using general principles of contact geometry and topology
outlined in our book (Kholodenko 2013). By doing so new directions in
detecting of these new knotted structures \ are being suggested.\bigskip

\textit{2.1. \ Dynamically generated knotted and linked structures} \bigskip 
\textit{and Chern-Simons }

\ \ \ \ \ \ \ \textit{topological field theory} \bigskip

Recent studies of dynamical systems revealed many instances in which knotted
and linked structures had been dynamically generated (Birman and Williams
1983, Ghrist, Holmes and Sullivan 1997, Ghys 2007), detected and classified.
Already mentioned correspondence between mechanics and fluid\ mechanics,
mechanics and electromagnetism, mechanics and geometrical optics provides
needed assurance for existence of knotted and linked structures, say, in
elctromagnetism. This correspondence \ is unusual from the standpoint of
currently existing opinion in physics literature. Indeed, beginning from the
work by Witten (1989) it is widely accepted that only the non-Abelian
version of the Chern-Simons (C-S) functional should be used for description
of nontrivial knots and links. The word "nontrivial" \ suggests knots/links
other than unknots, Hopf links and torus-type knots. The C-S functional
emerges naturally from the theory of pure Yang-Mills (Y-M) fields. In fact,
thanks to work of Floer (Donaldson 2002) it is possible to replace the
nonperturbative treatment of 4-dimensional Y-M gauge theory by the analogous
treatment of 3+1 gauge theory\ in which the 4 dimensional Y-M functional is
being replaced by the 3-dimensional C-S functional. This is in contrast with
some papers in physics literature in which both the Y-M and C-S functionals
\ are present in the initial action functional. Thus, because of this
replacement, when treated nonperturbatively, theory of pure Y-M fields
becomes topological. The C-S functional was already used for description of
knotted dynamical flows by Vejovsky and Freyer (1994). Because of this, the
question arises: Can general theory of non-Abelian gauge fields be used for
description of the Abelian version of these fields? Both Verjovsky and
Freyer (1994 ) and Trautman (1977) provided affirmative answer to this
question. Since the nonperturbative Y-M theory is topological, this then
implies that \ its Abelian version is also topological and, therefore,
should be capable of describing some knots/links. Trautman did not discuss
knots or links in \ his paper. Instead, he noticed that the Abelian (that is
of U(1)-type)) Y-M connection describing the Dirac monopole is solution of
the Maxwell equations very much like the SU(2)-type connection is the
solution of the non-Abelian Y-M. He also concluded that the analogous
treatment should be applicable to the gauge theory of gravity so that the
same computational protocol should yield gravitational instantons. In the
light of what follows, this is compatible with the statement that all gauge
field theories (Maxwell, Y-M and gravity) should contain solutions
describing knots/links. In this paper we shall not discuss gravity-related
topics. Interested reader should be able to find relevant information in
(Kholodenko 2011). \bigskip \bigskip

\textit{1.3. Connection with works by Ranada\bigskip }

By restricting \ ourself to the Maxwell and Y-M cases our treatment of
instantons \ in this work (which will be called part I) and its companion
(which will be called part II)\ formally differs (but equivalent) from that
developed by Trautman. The difference is caused by our desire to discuss
from the instanton perspective results of influential paper by Ranada (1989)
written much later.\ It had been in use in the\ majority of theoretical and
experimental works aimed at describing and detecting knotted beams of light.
The latest reference (Keida et al 2013) summarizes the latest efforts in
this direction. Ranada's original paper as well as many other subsequently
written either by him or with collaborators develop formalism without any
reference to (connection with) the non-Abelian theory of Y-M fields. As is
well known, due to their nonlinearity, the Y-M fields are much harder to
study than the Maxwellian fields. \ Numerous attempts to treat pure Y-M
fields by the saddle point methods resulted in negative answer to the
question about the existence of localized solutions (lumps or solitons) in
Minkowski space-time. In (Kholodenko 2011) a list of representative works of
various degree of rigor is provided in which this negative answer was
obtained. Absence of localized solutions makes it impossible to obtain
stable "knotty solutions" of the Y-M equations in Minkowski space-time. The
same conclusion holds for the Abelian (Maxwellian) gauge fields as it was
recently demonstrated in (Chubukalo et al 2010).

At the same time, in the Euclidean space the Y-M fields do have nontrivial
saddle point-type solutions \ known as instantons and monopoles. Ranada's
knots/links were designed to exist in the Minkowski space-time. Contrary to
the recent claims made in physics literature they cannot be immediately
compared with the electromagnetic instantons obtained by Trautman (Trautman
1977). In this work we demonstrate \ that, when properly interpreted,
Ranada's knots/links do have a chance to exist. \bigskip

\textit{1.4. Organization of the rest of this paper \bigskip }

As it was already stated, our work is made of two parts. The first part is
meant to reinterpret known (in physics literature) results for knots and
links in the Abelian gauge fields \ in terms of \ formalism of contact
geometry/topology. In doing so some new results are obtained \ to be listed
below. The second part is more technical and requires some in depth
knowledge of contact geometry/topology. It is being hoped that our readers
will consult whenever they are in doubt physics oriented monograph by
Kholodenko (2013) and, purely mathematically oriented monograph by Geiges
(2008).

Since, to our knowledge, Floer's work remains outside the scope of the
mainstream physics literature \ we provide in section 2 the self-contained
introduction to Floer's ideas. Section 3 contains some ramifications of
general results presented in section 2 aimed at reobtaining with help of
Floer's methodology designed for Y-M fields the Arnol'd inequality
extensively used in the book by Arnol'd and Khesin (1998) on topological
methods in hydrodynamics. Detailed study of this inequality is given in
section 4 in the context of hydrodynamics and electromagnetism. In this
section we begin to exploit the content of this inequality from the
perspective of contact geometry and topology. This allows us to recover the
major results of Ranada (1989,1992) and to reinterpret them
nontraditionally. Obtained results are aimed at preparing our readers for
extension of \ Ranada's results as well as those by (Keida et al 2013 and
Dennis et all 2010) to be discussed in part II. Part I concludes with
section 5 providing a summary and the list of tasks to be developed and
completed in the future publications.\bigskip

\textbf{2. Basics of Floer-style treatment of Yang-Mills instantons.
\bigskip }

\textit{2.1. \ Some basic facts about instantons \bigskip }

We begin our exposition of Floer's ideas by introducing some notations.
Following (Dubrovin et al 1984, Frankel 1997) it is sufficient to begin with
the Abelian (electromagnetic) case where the vector fields \textbf{E }and 
\textbf{B }\ represent various components of the second rank skew-symmetric
(electromagnetic) tensor $F_{ik}$ . It is\ determined at each point of, say,
Minkowski space-time (of signature 1,-1,-1,-1) as 
\begin{equation}
E_{\alpha }=F_{0\alpha },\alpha
=1,2,3.-B^{1}=F_{23},B^{2}=F_{13},-B^{3}=F_{12}  \tag{2.1a}
\end{equation}%
In terms of these notations, the 2-form $\mathbf{F}$\ is defined in a usual
way via 
\begin{equation}
\mathbf{F}=\frac{1}{2}F_{ij}dx^{i}\wedge dx^{j}=\tsum\limits_{\alpha
}E_{\alpha }dx^{0}\wedge dx^{\alpha }-B^{1}dx^{2}\wedge
dx^{3}+B^{2}dx^{1}\wedge dx^{3}-B^{3}dx^{1}\wedge dx^{2}  \tag{2.1b}
\end{equation}%
and its dual $\ast \mathbf{F}$ via 
\begin{equation}
\ast \mathbf{F}=-\tsum\limits_{\alpha }B_{\alpha }dx^{0}\wedge dx^{\alpha
}-E_{1}dx^{2}\wedge dx^{3}+E_{2}dx^{1}\wedge dx^{3}-E_{3}dx^{1}\wedge dx^{2}.
\tag{2.1c}
\end{equation}%
From here, it follows that ($\ast \mathbf{F}\mathcal{)}_{ij}=\frac{1}{2}%
\varepsilon _{ijlm}F^{lm},F^{lm}=g^{lp}g^{mq}F_{pq}$ with Minkowski metric
tensor $g_{ij},\varepsilon ^{0123}$ $=1$, and $g^{ij}=\left( g_{ij}\right)
^{-1}.$ With help of these definitions we obtain: $F^{0\alpha }=-F_{0\alpha
},$ and $F^{\alpha \beta }=F_{\alpha \beta },\alpha ,\beta =1,2,3.$ Thus, $%
\ast \left( \ast \mathbf{F}\right) =-\mathbf{F}$. That is the square of the
Hodge star operator in Minkowski space-time is equal to $-1$.

The antisymmetric tensor $F_{pq}$ has 6 independent components. These
components\ can be treated as components of some vector in \textbf{R}$^{6}.$
Because of this, \ introduce complex variable via $\tilde{z}=a+ib$ then, let 
$\tilde{z}\mathbf{F}\mathcal{\equiv }a\mathbf{F}+b\ast \mathbf{F}$. Under
such an identification $i^{2}\mathbf{F}=-\mathbf{F}$. \ This fact allows us
to replace \textbf{R}$^{6}$ by \textbf{C}$^{3}$ and to identify the Hodge
star operation with that of complex multiplication. In \textbf{C}$^{3}$ it
is possible to introduce complex coordinates $z^{\alpha }$ via 
\begin{equation}
z^{\alpha }=E_{\alpha }+iB^{\alpha },\alpha =1,2,3.  \tag{2.2}
\end{equation}%
Such complex vector is known in literature as Riemann-Silberstein (RS)
vector. The Maxwellian theory of electromagnetism was extensively discussed
in terms of the RS vector in recent review paper (Bialynicki-Birula 2013).
By design, uses of this vector are limited to space-times of Minkowski \
signature. \ In such a case, one typically introduces the quadratic form
(Dubrovin 1984) $(\mathbf{E}+i\mathbf{B})^{2}$ =$\tsum\limits_{\alpha
=1}^{3} $ $\left( z^{\alpha }\right) ^{2}$ enabling us to classify various
possibilities for vector fields $\mathbf{E}$ and $\mathbf{H}$. It is
believed (Kedia 2013), that the electromagnetic knots can be formed only by
null fields (read, however, Appendix D). These are determined by the
condition%
\begin{equation}
\tsum\limits_{\alpha =1}^{3}\left( z^{\alpha }\right) ^{2}=0.  \tag{2.3}
\end{equation}%
This condition is formally satisfied if $\mathbf{E}+i\mathbf{B=}0$ which is
mathematical statement of the anti-self-duality in spacetimes of Minkowski
signature (Mason and Woodhouse 1996, Chubukalo et al 2010). Clearly, for
this to make sense we have to treat both \textbf{E} and \textbf{B }as
complex numbers thus contradicting the initial assumption made in eq.(2.2).
Nevertheless, for the sake of results presented below, it is helpful to know
that historically, the concept of (anti)-self-duality is associated with the
Y-M instantons "living" in Euclidean space. This concept can be used in
space-times of Minkowski signature also (Mason and Woodhouse 1996, Nash and
Sen 1983, Chubukalo et al 2010). As it was noticed in the Introduction, in
such space-times the Y-M action functional \ calculated on (anti)-self-dual
fields vanishes. Evidently, this result cannot be immediately applied to the
RS vector. Instead, we have to look \ for another\ option, e.g.%
\begin{equation}
\mathbf{E}^{2}-\mathbf{H}^{2}+2i\mathbf{E}\cdot \mathbf{H}=0,  \tag{2.4}
\end{equation}%
where the symbol $\cdot $ denotes the Euclidean space scalar product. From
here we obtain two conditions: $\left\vert \mathbf{E}\right\vert =\left\vert 
\mathbf{H}\right\vert $ and $\mathbf{E}\cdot \mathbf{H=}0\mathbf{.}$ These
two properties, by design, characterize the \textit{null fields}. \ Since 
\textit{any} $\mathbf{F}$\textbf{\ }can be decomposed\textbf{\ }as $\mathbf{F%
}=\mathbf{F}^{+}+\mathbf{F}^{-}$ where $\mathbf{F}^{+}=\frac{1}{2}(\mathbf{F}%
+\ast \mathbf{F})$ and $\mathbf{F}^{-}=\frac{1}{2}(\mathbf{F}-\ast \mathbf{F}%
),$ \ it is clear, that the above null fields can be represented via linear
combination of dual and anti-self-dual fields. This option is suggested in
(Chubukalo 2010) without any reference to Ranada's results. It happens,
however, that Ranada's results indeed fall into exactly this category as
will be explained later in the text.

The action functional $S$ for both the Y-M and Maxwellian fields can be
written (up to a constant factor) as (Frankel 1996) 
\begin{equation}
S[\mathbf{F}]=-tr\tint\limits_{M}(\mathbf{F}\wedge \ast \mathbf{F})\equiv
\left\Vert \mathbf{F}\right\Vert ^{2}  \tag{2.5}
\end{equation}%
Following \ Floer (Donaldson 2002) it is convenient to design the 4-manifold 
$\mathcal{M}$ as direct product of some 3-manifold $Y$ (in the simplest case
it is $S^{3})$ and time $\mathbf{R}$ , that is $\mathcal{M}$ $=Y\times 
\mathbf{R}$. The sign "-" in front of the integral in eq.(2.5) is written in
accord with the Abelian case (Landau and Lifshitz 1975) treated in spaces of
Minkowski signature.

From the discussion related to the RS vector it follows that in spacetimes
of Minkowski signature \ the (anti)- self-duality equation/condition is
given by 
\begin{equation}
\ast \mathbf{F}\mathcal{=\pm }i\mathbf{F}.  \tag{2.6}
\end{equation}%
This result makes sense for both Abelian and non-Abelian gauge fields (Nash
and Sen 1983). Switching to spaces with Euclidean signature results in
replacing eq.(2.6) by 
\begin{equation}
\ast \mathbf{F}\mathcal{=\pm }\mathbf{F.}  \tag{2.7}
\end{equation}%
\ In the absence of sources the first pair of Maxwell's equations in
Minkowski space-time are given by 
\begin{equation}
\text{div}\mathbf{B}=0,\text{curl}\mathbf{E}+\frac{\partial \mathbf{B}}{%
\partial t}=0.  \tag{2.8a}
\end{equation}%
These are equivalent to the condition: $d\mathbf{F}=0,$ that is to the
Bianchi identity. \ The second pair \ of Maxwell's equations are given by%
\begin{equation}
\text{div}\mathbf{E}=0,\text{curl}\mathbf{B}-\frac{\partial \mathbf{E}}{%
\partial t}=0.  \tag{2.8b}
\end{equation}%
These are equivalent to the equation $d\ast \mathbf{F}=0.$ Both $d\mathbf{F}%
=0$ and \ $d\ast \mathbf{F}=0$ are obtainable from the action functional,
eq. (2.5). These equations look formally the same for both the Y-M and
Maxwellian fields. \ The anti-self-duality equation/condition in the
Minkowski space-time is given by 
\begin{equation}
\mathbf{B}=-i\mathbf{E}  \tag{2.9a}
\end{equation}%
while in the Euclidean space it is given by 
\begin{equation}
\mathbf{B}=-\mathbf{E}.  \tag{2.9b}
\end{equation}%
\bigskip \bigskip

\textit{2.2. \ Connections with the paper by Ranada (1989)\bigskip .
Emergence of dyons}

Formally, the last result, eq.(2.9b), is just eq.(13) of Ranada's paper
(Ranada 1989). In fact, Ranada uses both the duality, that is $\mathbf{B}=%
\mathbf{E,}$ and the anti-self duality conditions. Such an interpretation of
his results is superficial though. It is superficial because eq.(2.9a) is
actually written by Ranada as $\mathbf{B(}\theta \mathbf{)}=-\mathbf{E(}\phi 
\mathbf{)}$ and , accordingly, $\mathbf{B(}\phi \mathbf{)}=\mathbf{E(}\theta 
\mathbf{).}$ Strictly speaking, these are not (anti)-self-dual fields since
for such fields the arguments in the above equations should be the same. How
then one should understand these equations? \ Well, using both of the above
equations leads to the conditions: $\mathbf{B(}\theta \mathbf{)\cdot E(}%
\theta \mathbf{)=-B(}\phi \mathbf{)\cdot E(}\phi \mathbf{)}$ and $\mathbf{B}%
^{2}\mathbf{(}\theta \mathbf{)=E}^{2}\mathbf{(}\phi \mathbf{),B}^{2}\mathbf{(%
}\phi \mathbf{)=E}^{2}\mathbf{(}\theta \mathbf{).}$ Should the arguments in
all these equations be the same, then these equations would \ indeed
describe the null fields. But they are not the same! \ Again, how then one
should understand these equations?

To inject some physics into these thus far formal results, it is helpful to
notice that Maxwell's eq.s(2.8) will remain invariant under formal
replacement: $\mathbf{E}\rightarrow -\mathbf{B}$, $\mathbf{B}\rightarrow 
\mathbf{E}.$ These are the electric-magnetic duality transformations
discovered by Heviside in 1893 as described in the paper by Mignaco (Mignaco
2001). Maxwell's equations with charges and currents loose this type of
invariance, unless the magnetic monopoles are present. Recall, that
according to Trautman the U(1) fiber bundle connection corresponding to the
magnetic pole is nontrivial. This connection solves Maxwell's equations
while the SU(2) connection solves the non-Abelian Y-M equations yielding the
instanton and monopole solutions. In his papers (Ranada 1989, 1992) Ranada
indeed obtains the connection and curvature known for the Dirac monopole.
But, in addition, by assuming the electric-magnetic duality he treats the
electric charges as electric-type monopoles too. As result, if \ Trautman's
electromagnetic field is the field originating from the Dirac monopole,
Ranada's electromagnetic field is the field originating from the Abelian
dyon, that is from the hypothetical particle which is carrying the magnetic
and electric charges simultaneously. Since neither Ranada and his
collaborators nor those who used his results had recognized Ranada's
construction of electromagnetic field as that attributed to the Abelian
dyons (Pakman 2000, Negi and Dehnen 2011), in appendix A we provide some
very basic information on dyons. The dyonic interpretation of Ranada's
equations $\mathbf{B(}\theta \mathbf{)}=-\mathbf{E(}\phi \mathbf{)}$ and $%
\mathbf{B(}\phi \mathbf{)}=\mathbf{E(}\theta \mathbf{)}$ converts them into
(anti)-self-duality conditions in Euclidean space. At the same time, since
the scalar product in eq.(2.4) is defined in Euclidean space (having in mind
3+1 decomposition \ of Minkowski space-time) Ranada's equations, in fact,
define the null fields in Minkowski space-time in the sense we just had
described. \ The 3+1 decomposition used by Floer allows us to understand
this apparent peculiarity without difficulty. This is explained in appendix
B.\ At this point the attentive reader would object to our formal dyonic
interpretation of Ranada's results\vspace{-0.66in} since Ranada's papers
describe only

\bigskip

\bigskip

\bigskip

\bigskip

\bigskip

\QTP{Body Math}
the electromagnetic fields \textsl{without charges. }This objection can be
removed, however, as follows. For the sake of argument, consider the case of
monopoles first. In our book (Kholodenko 2013) we \ explained in great
detail that the (Dirac) monopole can be recreated with help of
superconducting ring in which the superconducting current flows. The
magnetic field outside the ring does not have sources and thus forms a
complementary ring which \ is linked with the magnetic field \ located on
the surface of the superconducting ring. Thus, we are dealing with the
Hopf-type interlocked rings. As result, the magnetic field coming from such
ring system is \ indistinguishable from the field originated from the Dirac
monopole. This requires some proof which is given in (Kholodenko 2013),
Chapter 3, section 3.5. In Ranada's case the Dirac monopole is being
modelled by the system of interlocked magnetic rings. From here, there is no
need for the magnetic charge! Because of the electric-magnetic duality,
following Ranada, we have to add to these two magnetic rings another two
interlocked electric fields. Such interlocked electric rings serve to
replace the electric charge. Thus, both electric and magnetic charges can be
described in terms of the corresponding interlocked Hopfian rings. Since the
field (electric or magnetic) originating from such Hopfian rings is
indistinguisheable from that coming from "true" electric or magnetic
charges, the presence of such topological formations in electromagnetic
field spares us from the necessity to have actual charges. \ In view of such
an interpretation, the situation in the Abelian case now parallels that in
the non-Abelian case (Manton and Suttcliffe 2007) where monopoles and dyons
are created directly from the sourceless non-Abelian Y-M fields. In the
Abelian case in order to obtain a dyon \textit{without actually having
electric or magnetic charge }it is sufficient to construct a system made of
two interlocked magnetic rings to which two interlocked electrical rings
should be added. Such a system is being described in terms of the
electromagnetic field designed by Ranada in 1989. This interpretation of
Ranada's results leads to far reaching consequences. They are going to be
briefly discussed in section 5.\medskip

\bigskip

\bigskip

\textit{2.3. \ \ Back to instantons \medskip \vspace{-0.66in}\bigskip
\bigskip }

\bigskip

\bigskip

\bigskip

To develop needed formalism we have to finish with general topics related to
instantons. Some auxiliary information needed for connecting Floer's
arguments with those known in physics literature is collected in Appendix A.
\ From this appendix it follows that in Euclidean space the action
functional, eq.(2.5), can be written as 
\begin{equation}
S=\frac{1}{2}\tint\limits_{\mathcal{M}_{E}}dv(4)[\mathbf{E}^{2}+\mathbf{B}%
^{2}],  \tag{2.10a}
\end{equation}%
where $\mathcal{M}_{E}$ stands for 4-manifold of Euclidean signature. Here $%
dv(4)$ stands for the volume element for such manifold. In the space-time of
Minkowski signature the same action $S$ is known (Landau and Lifshitz 1975)
as 
\begin{equation}
S=\frac{1}{2}\tint\limits_{\mathcal{M}}dv(3,1)[\mathbf{E}^{2}-\mathbf{B}%
^{2}].  \tag{2.10b}
\end{equation}%
It is helpful to add few details to these results. First, we notice that the
action is determined with accuracy up to some space/time derivative of some
scalar function of $\mathbf{E}$ and $\mathbf{B}$ vector fields. The fact
that this derivative can be dropped is the result of imposed space-time
boundary conditions which should be specified both in Minkowski and
Euclidean spaces. Second, it is helpful to recall that if $\mathcal{M}$ $%
=Y\times \mathbf{R}$ in eq.(2.10b), then (up to a constant) 
\begin{equation}
\mathcal{E}=\frac{1}{2}\tint\limits_{Y}dv(3)[\mathbf{E}^{2}+\mathbf{B}^{2}] 
\tag{2.11}
\end{equation}%
is the genuine energy density of electromagnetic (Landau and Lifshitz 1975)
or Yang-Mills Frankel 1997) fields. By comparing eq.(2.10a) with eq.(2.11) \
it is evident that in both cases the energy density is the same. This
observation will be frequently used in both parts of our work.

To proceed with instantons, in view of eq.(2.5) we also need to introduce
the 2nd Chern number (or instanton topological charge) $C_{2}$ which (up to
a constant factor) is given by 
\begin{equation}
C_{2}\simeq tr\tint\limits_{\mathcal{M}}(\mathbf{F}\wedge \mathbf{F}). 
\tag{2.12}
\end{equation}%
The constant factor is determined by the gauge group which is in use. Since $%
tr\tint\limits_{\mathcal{M}}(\ast \mathbf{F}\wedge \mathbf{F}%
)=tr\tint\limits_{\mathcal{M}}(\mathbf{F}\wedge \ast \mathbf{F})),$ we also
obtain,%
\begin{equation}
tr\tint\limits_{\mathcal{M}}(\ast \mathbf{F}\wedge \ast \mathbf{F}%
)=tr\tint\limits_{\mathcal{M}}(\mathbf{F}\wedge \mathbf{F}).  \tag{2.13}
\end{equation}%
Using this result we rewrite the action in eq.(2.5) as 
\begin{equation}
S[\mathbf{F}]=-\frac{1}{2}tr\tint\limits_{\mathcal{M}}(\mathbf{F}+\ast 
\mathbf{F}\mathcal{)}\wedge (\mathbf{F}\mathcal{+}\ast \mathbf{F}%
)+tr\tint\limits_{\mathcal{M}}(\mathbf{F}\wedge \mathbf{F})\geq
tr\tint\limits_{\mathcal{M}}(\mathbf{F}\wedge \mathbf{F})  \tag{2.14}
\end{equation}%
with the equality achieved for the anti-self-dual solutions:\ $\ast \mathbf{F%
}\mathcal{=-}\mathbf{F}$. \ Since it is possible to write the analogous
inequality for the dual fields too (Manton and Sutcliffe 2007), use of this
inequality is not making the anti-self-dual fields more special than the
dual ones. Nevertheless, in mathematics literature, e.g. read (Donaldson and
Kronheimer 1990), pages 38-39, 43-47, there is an explaination of the fact
that in instanton calculations it is mathematically incorrect to use
solutions of both duality, that is $\mathbf{F}^{-}=0,$ and
anti-self-duality, that is $\mathbf{F}^{+}=0,$ equations simultaneously.
This restriction is associated with the choice of complex structure on the
underlying manifold. Once the complex structure is selected, it is
mathematically incorrect to switch to another structure. Since the null
fields require both duality and anti-self-duality solutions for their
realization, this means that Trautman's monopole results (Trautman 1977) are
in accord with the existing mathematical restrictions while Ranada's dyonic
solution (Ranada 1989,1992) still requires justification which will be
provided below.

Next, following Frankel (1997) we write (symbolically) for curvature $%
\mathbf{F}=d\mathbf{A}+\mathbf{A}\wedge \mathbf{A}.$ Therefore, for any
curvature 2-form matrix \ which is defined for $any$ vector bundle over $any$
manifold of $any$ dimension the following chain of equalities holds: 
\begin{eqnarray*}
\mathbf{F}\wedge \mathbf{F} &\mathcal{=}&\mathcal{(}d\mathbf{A}+\mathbf{A}%
\wedge \mathbf{A})\wedge \mathcal{(}d\mathbf{A}+\mathbf{A}\wedge \mathbf{A})=
\\
&&d\mathbf{A}\wedge d\mathbf{A}+d\mathbf{A}\wedge \mathbf{A}\wedge \mathbf{A}%
+\mathbf{A}\wedge \mathbf{A}\wedge d\mathbf{A}+\mathbf{A}\wedge \mathbf{A}%
\wedge \mathbf{A}\wedge \mathbf{A}.
\end{eqnarray*}%
By applying the trace operation to the above expression and by keeping in
mind that $tr(\mathbf{A}\wedge \mathbf{A}\wedge \mathbf{A}\wedge \mathbf{A}%
)=0$ and $tr(d\mathbf{A}\wedge \mathbf{A}\wedge \mathbf{A})=tr(\mathbf{A}%
\wedge \mathbf{A}\wedge d\mathbf{A}),etc.$ we eventually arrive at the
crucial identity 
\begin{equation}
tr\left( \mathbf{F}\wedge \mathbf{F}\right) =dtr(\mathbf{A}\wedge d\mathbf{A}%
+\frac{2}{3}\mathbf{A}\wedge \mathbf{A}\wedge \mathbf{A}).  \tag{2.15}
\end{equation}%
This identity, when being used in eq.(2.14), leads to the following result: 
\begin{eqnarray}
C_{2} &\simeq &tr\tint\limits_{\mathcal{M}}(\mathbf{F}\wedge \mathbf{F}%
)=\tint\limits_{\mathcal{M}}trd(\mathbf{A}\wedge d\mathbf{A}+\frac{2}{3}%
\mathbf{A}\wedge \mathbf{A}\wedge \mathbf{A})  \notag \\
&=&\tint\limits_{\partial \mathcal{M}}tr(\mathbf{A}\wedge d\mathbf{A}+\frac{2%
}{3}\mathbf{A}\wedge \mathbf{A}\wedge \mathbf{A})\risingdotseq CS(\mathbf{A}%
).  \TCItag{2.16}
\end{eqnarray}%
Here $CS(\mathbf{A})$ stands for the "Chern-Simons functional" and $%
\risingdotseq $ means "up to a constant" (determined by the gauge group
being \ in use). This derivation assumes that $\mathcal{M}$ is the manifold
with boundary. It is not limited to 3 dimensions. The CS-like functionals
exist for manifolds of any odd dimensions and, accordingly, the Y-M-like
fields can be defined on even dimensional manifolds of dimensionality one
higher.

The results just obtained imply 
\begin{equation}
S[\mathbf{F}]\geq \varkappa CS(\mathbf{A}),  \tag{2.17}
\end{equation}%
where $\varkappa $ is some gauge group-dependent constant factor. Therefore,
it follows that the minima of $S[\mathbf{F}]$ are determined by the minima
of $CS(\mathbf{A}).$ These are given by the zero curvature condition (valid
at the boundary $\partial \mathcal{M)}$ 
\begin{equation}
\frac{\delta CS(\mathbf{A})}{\delta \mathbf{A}}=\mathbf{F[A]}=d\mathbf{A}+%
\mathbf{A}\wedge \mathbf{A=}0  \tag{2.18}
\end{equation}%
obtainable by minimization of $CS(\mathbf{A})$. Details of minimization
calculations can be found, for example, in (Jost $\ 2008)$. The obtained
result leads to the apparent contradiction. \ Indeed, the $CS(\mathbf{A})$
functional is not invariant with respect to gauge transformations (Donaldson
2002, Manton and Sutcliffe 2007). This causes $CS(\mathbf{A})$ to be
determined only with accuracy up to some integer. In addition, the field 
\textbf{A} \ used in $S[\mathbf{F}]$ depends upon four independent
(space-time) variables while the field \textbf{A} in $CS(\mathbf{A})$ \
apparently depends upon 3 variables only. How these difficulties are being
resolved in physics literature we shall not discuss in what follows.
Instead, we are going to use Floer's ideas, To facilitate understanding of
Floer's work\footnote{%
That is that part of Floer's works which is needed for the purposes of this
paper.}, we recommend reading of appendix A \ inspired by Floer's ideas
prior to reading the rest of this section.

\bigskip

\textit{2.4. \ \ Basics of Floer's input to the Y-M instantons}

\textit{\bigskip }

Our exposition of Floer's ideas is based in part on results \ from the
Donaldson's monograph (Donaldson 2002). From this work it follows that it is
very helpful to discuss the minimization problem for Y-M fields in the
context of classical mechanics. \ For this purpose, it is necessary to
reformulate known in physics mechanical formalism somewhat.

Thus, let $q_{i}(t)$ be the i-th component of coordinate describing the
trajectory of a particle of unit mass in the potential $V$. Then, Newton's
equations of motion are given by 
\begin{equation}
\frac{d^{2}}{dt^{2}}q_{i}=-\nabla _{i}V.  \tag{2.19}
\end{equation}%
These equations are obtainable by variation of the action functional $%
S[q(t)] $ defined by 
\begin{equation}
S[q(t)]=\tint\limits_{0}^{t}d\tau \lbrack \frac{1}{2}\dot{q}^{2}-V(q)]. 
\tag{2.20}
\end{equation}%
The unexpected twist in this well known protocol originates from the
following observation. Suppose that there is some function $\sigma (q)$ such
that the potential $V$ can be represented as 
\begin{equation}
V=-\frac{1}{2}\left\vert \nabla \sigma \right\vert ^{2}.  \tag{2.21}
\end{equation}%
Then, the first order equation 
\begin{equation}
\dot{q}_{i}=\nabla _{i}\sigma  \tag{2.22}
\end{equation}%
will have the same solutions as Newton's \ eq.(2.19). \ 

To prove that this is indeed the case, let us consider \ the following chain
of equalities:%
\begin{equation}
\frac{d}{dt}\dot{q}_{i}=\frac{d}{dt}(\nabla _{i}\sigma )=\tsum\limits_{j}%
\frac{\partial }{\partial q_{j}}(\frac{\partial \sigma }{\partial q_{i}})%
\frac{dq_{j}}{dt}  \tag{2.23}
\end{equation}%
By combining this result with eq.(2.22) we obtain: 
\begin{equation}
\frac{d}{dt}\dot{q}_{i}=\tsum\limits_{j}\frac{\partial ^{2}\sigma }{\partial
q_{i}\partial q_{j}}\frac{\partial \sigma }{\partial q_{j}}.  \tag{2.24}
\end{equation}%
At the same time, 
\begin{equation}
\frac{\partial }{\partial q_{i}}V=-\frac{1}{2}\frac{\partial }{\partial q_{i}%
}\tsum\limits_{j}\left( \frac{\partial \sigma }{\partial q_{j}}\right)
^{2}=-\tsum\limits_{j}\frac{\partial ^{2}\sigma }{\partial q_{i}\partial
q_{j}}\frac{\partial \sigma }{\partial q_{j}}.  \tag{2.25}
\end{equation}%
Using this result in (2.19) we reobtain (2.24) as required. This fact allows
us to rewrite the action functional $S[q(t)]$ in the form analogous to
eq.(A.5) for the Y-M fields. Indeed, we obtain,%
\begin{eqnarray}
S[q(t)] &=&\tint\limits_{0}^{t}d\tau \lbrack \frac{1}{2}\dot{q}%
^{2}-V(q)]=\tint\limits_{0}^{t}d\tau \frac{1}{2}[\dot{q}^{2}+\left\vert
\nabla \sigma \right\vert ^{2}]  \notag \\
&=&\tint\limits_{0}^{t}d\tau \tsum\limits_{i}[\frac{1}{2}\{[\dot{q}%
_{i}^{2}-\nabla _{i}\sigma ]^{2}\}+\frac{\partial \sigma }{\partial q_{i}}%
\frac{dq_{i}}{d\tau }]  \notag \\
&=&\tint\limits_{0}^{t}d\tau \lbrack \frac{1}{2}\{\left\vert \dot{q}-\nabla
\sigma \right\vert ^{2}+\frac{d\sigma }{d\tau }].  \TCItag{2.26a}
\end{eqnarray}%
Thus, when $\dot{q}_{i}=\nabla _{i}\sigma $ we obtain, 
\begin{equation}
S[q(t)]=\sigma (\tilde{q}(t))-\sigma (q(0))  \tag{2.26b}
\end{equation}%
where $\tilde{q}(t)$ is the solution of eq(2.22) at time $t$. Therefore, 
\begin{equation}
S[q(t)]\geq \lbrack \sigma (\tilde{q}(t))-\sigma (q(0))].  \tag{2.27}
\end{equation}%
Now by rewriting eq.(A.5) as%
\begin{equation}
S[\mathbf{A}]=\frac{1}{2}\tint\limits_{-\infty }^{\infty
}dt\tint\limits_{Y}dv[\mathbf{\dot{A}}^{2}+\mathbf{B}^{2}],  \tag{2.28}
\end{equation}%
since $\mathbf{\nabla }\times \mathbf{A=B}$ and by comparing it against
eq.(2.26a) we obtain the following correspondences 
\begin{equation}
q_{i}\rightleftarrows A_{i}\text{ , }\frac{\partial }{\partial q_{i}}%
\rightleftarrows \frac{\partial }{\partial A_{i}},\text{ }\left\vert \nabla
\sigma \right\vert ^{2}=\nabla \sigma \cdot \nabla \sigma \rightleftarrows 
\mathbf{B}\cdot \mathbf{B}  \tag{2.29}
\end{equation}%
Furthermore, using \ eq.(2.18) we can write as well 
\begin{equation}
\nabla \sigma \cdot \nabla \sigma \rightleftarrows \tsum\limits_{i}\frac{%
\delta CS(\mathbf{A})}{\delta A_{i}}\frac{\delta CS(\mathbf{A})}{\delta A_{i}%
}  \tag{2.30}
\end{equation}%
so that 
\begin{equation*}
\sigma \rightleftarrows CS(\mathbf{A}).
\end{equation*}%
This requires us to prove that 
\begin{equation}
\frac{\delta CS(\mathbf{A})}{\delta A_{i}}=\pm B_{i}.  \tag{2.31}
\end{equation}%
This task is accomplished in appendix B. But, if eq.(2.31) is correct then,
using eq.s(2.22) and (2.31) we obtain as well%
\begin{equation}
\frac{\partial A_{i}}{\partial t}=\pm B_{i}.  \tag{2.32}
\end{equation}%
To decide which sign to choose we have to look at the interpretation of this
result in the Abelian case. For such a case we should use the
anti-self-duality, eq(2.9), and, from appendix A, the fact that $\mathbf{E}=-%
\frac{\partial }{\partial t}\mathbf{A.}$ This then helps us to reach the
conclusion that we must choose the sign "+" in the above anti-self-duality
equation. Naturally, it is equivalent to eq.(2.9) as required. In view of
eq.(2.31) it is convenient to rewrite eq.(2.32) in the form of\ the gradient
flow equation 
\begin{equation}
\frac{\partial A_{i}}{\partial t}=\frac{\delta CS(\mathbf{A})}{\delta A_{i}}.
\tag{2.33}
\end{equation}%
Since eq.s (2.22) and (2.33) are equivalent, we can interpret both of them
as describing critical dynamics of some kind of statistical system. More on
this can be found in (Kholodenko 2008)\footnote{%
Interested readers are encouraged also to consult (Oh 2013) and (Sreets
2007).}. Eq.(2.33) is nontrivial and leads to the extremely sophisticated
analysis done by Floer (1988). Fortunately, we do not need his analysis in
this work. Instead, it is helpful now to reconsider once again eq.(2.28). We
obtain,%
\begin{eqnarray}
S[\mathbf{A}] &=&\frac{1}{2}\tint\limits_{-\infty }^{\infty
}dt\tint\limits_{Y}dv(3)[\mathbf{\dot{A}}^{2}+\mathbf{B}^{2}]=\frac{1}{2}%
\tint\limits_{-\infty }^{\infty }dt\tint\limits_{Y}dv(3)[\mathbf{E}^{2}+%
\mathbf{B}^{2}]  \notag \\
&=&\tint\limits_{-\infty }^{\infty }dt\tint\limits_{Y}dv(3)\{\frac{1}{2}%
\left\vert \mathbf{E}+\mathbf{B}\right\vert ^{2}-\mathbf{E}\cdot \mathbf{B\}.%
}  \TCItag{2.34}
\end{eqnarray}%
Deser and Teitelboim (1976) noticed that the $\mathbf{E}\cdot \mathbf{B}$
term in the above integral can be represented as the total divergence, i.e. $%
\partial _{\mu }C^{\mu \text{ }}$where $C^{\mu }=\varepsilon ^{\mu \nu \rho
\sigma }A_{\nu }\partial _{\rho }A_{\sigma }$ . Because of this, when \ the
anti-self-duality condition, eq(2.9.b), is fulfilled, we obtain: 
\begin{equation}
S[\mathbf{A}]=\tint\limits_{-\infty }^{\infty }dt\tint\limits_{Y}dv\mathbf{B}%
^{2}\geq CS(\mathbf{A}_{\infty })-CS(\mathbf{A}_{-\infty }).  \tag{2.35}
\end{equation}%
It makes physical sense to require $\mathbf{A}_{-\infty }=0$ so that $CS(%
\mathbf{A}_{-\infty })=0.$ This then leads us to the recovery of the already
known result, eq.(2.17). \ As plausible as it is, this fact is not
sufficient for the obtained results to be used further. Explanations are
provided in the next section.\medskip

\textbf{3. Ramifications\bigskip }

To move forward, we must take into account that the CS functional is not an
invariant of gauge transformations. To deal with this fact we need to
introduce several definitions. First, we would like to remind to our readers
that the $\mathbf{A}$-field (the connection) under gauge transformations is
transformed as 
\begin{equation}
\mathbf{A\rightarrow gAg-dgg}^{-1}\equiv \mathbf{g}(\mathbf{A})  \tag{3.1}
\end{equation}%
where $\mathbf{g}\in \mathcal{G}$ where $\mathcal{G}$ is the gauge group of
automorphisms of whose Lie algebra is determined by the known set of
generators $T_{\alpha }$. As it follows from the above definition, $\mathbf{g%
}$ is coordinate-dependent in general. Because of this, we have to operate
not with the space $\mathcal{A}$ of all connections \ but with its quotient $%
\mathcal{B}=\mathcal{A}/\mathcal{G}$ (roughly equivalent to the moduli
space). $\mathcal{B}$ is a smooth infinite-dimensional manifold\footnote{%
We skip the fact that $\mathcal{B}$ is Banach manifold whose definition is
given in (Donaldson 2002)}. \ The moduli space is the subset $\mathcal{R}$ $%
\subset $ $\mathcal{B}$ defined as the set of all flat connections via 
\begin{equation}
\mathcal{R}=\{[A]\in \mathcal{B\mid }\mathbf{F}[\mathbf{A}]=0\}.  \tag{3.2}
\end{equation}%
It is invariant with respect to the action of $\mathcal{G}.$ We can
temporarily fix the constant in the CS functional, eq.(2.15), if, following
Floer (1988), we define the CS functional as%
\begin{equation}
\frac{1}{2}\tint\limits_{\partial M}tr(\mathbf{A}\wedge d\mathbf{A}+\frac{2}{%
3}\mathbf{A}\wedge \mathbf{A}\wedge \mathbf{A})=CS(\mathbf{A}).  \tag{3.3}
\end{equation}%
Then, gauge transformations defined by eq.(3.1) \ when applied to $CS(%
\mathbf{A})$ produces%
\begin{equation}
CS(\mathbf{g}(\mathbf{A}))=CS(\mathbf{A})+2\pi \deg (\mathbf{g}),  \tag{3.4}
\end{equation}%
where deg(\textbf{g}) is some integer. It is the degree of a map between
3-dimensional closed manifolds. Based on this result, Floer concludes that $%
CS(\mathbf{A})$ is well defined on the quotient 
\begin{equation}
\mathcal{\tilde{B}=A}/\{\mathbf{g\in }\mathcal{G\mid }\deg (\mathbf{g})=0\} 
\tag{3.5}
\end{equation}%
\ The gradient flow, eq.(2.33), is well defined only on $\mathcal{\tilde{B}}$%
. Imposition of the constraint, eq.(3.5), is the major source of differences
in treatments of Y-M theory in physics and mathematics. In this work we
follow mathematician's path as stated already.

The condition, eq.(3.5), is easy to understand. Indeed, the minimization of
eq.(2.18) produces some flat connection $\mathbf{F[A]}=0.$ It belongs to the
space $\mathcal{R}$. This connection must be such that the action, eq.(2.3),
stays finite. For this reason and the fact that dynamics takes place in
Euclidean space, such connections are called instantons. In fact, these are
identical with the gauge equivalence classes, say, $a$ and $b,$ defined as
follows. Recall that instantons in the usual quantum mechanics travel from
one hump/lump/maximum (critical point $\equiv $ vac$\unit{u}$um) \ of the
inverted potential to another (critical point, that is another vacuum). The
maxima are critical points of the underlying manifold. This fact is used in
the Morse theory to recover some topological characteristics of the
underlying manifold. In the Y-M theory the maxima are determined by
eq.(2.18) and the CS functional is playing a role of the Morse function. In
such interpretation $a$ and $b$ are different (in general) equivalence
classes associated with different critical points of $\mathcal{R}.$
Furthermore, it is possible to decompose $\mathcal{R}$ into subspaces $%
\mathcal{M}(a,b)$ of instantons \ connecting $a$ and $b$ $\in \mathcal{R}.$
Thus, \ Floer extended the conventional Morse theory so that it becomes
capable of computing the topological properties of $\mathcal{R}$. Evidently,
if the restriction given by Eq.(3.5) is not imposed, then one cannot talk
about well defined classes $a$ and $b$.

Eq.(3.4) is the statement of periodicity of the CS functional. \ Following
Donaldson (2002) this periodicity can be equivalently stated as follows. For
the sake of argument we shall use the gauge group $SU(2)$. Then, 
\begin{equation*}
C_{2}=\frac{1}{8\pi ^{2}}tr\tint\limits_{[0,t]\times Y}(\mathbf{F}\wedge 
\mathbf{F})\equiv \theta (t).
\end{equation*}%
Accordingly,%
\begin{equation}
\theta (t_{2})-\theta (t_{1})=\frac{1}{8\pi ^{2}}tr\tint%
\limits_{[t_{1},t_{2}]\times Y}(\mathbf{F}\wedge \mathbf{F})\text{ }\func{mod%
}\mathbf{Z,}  \tag{3.6}
\end{equation}%
where \textbf{Z} denotes the ring of integers. In view of this equation and,
taking into account eq.(2.15), we obtain (for any gauge group!)%
\begin{equation}
tr\tint\limits_{[0,1]\times Y}(\mathbf{F}\wedge \mathbf{F}%
)=\tint\limits_{Y}tr(\mathbf{A}\wedge d\mathbf{A}+\frac{2}{3}\mathbf{A}%
\wedge \mathbf{A}\wedge \mathbf{A})  \tag{3.7}
\end{equation}%
This result should be understood as follows. Choose a connection \textbf{A}
on the principal $G$ bundle over 4-manifold $[0,1]\times Y=S^{1}\times Y$ in
such a way that for \{0\}$\times Y$ it is trivial while for \{1\}$\times Y$
it is not trivial. Then we have to consider the degree of the map $%
S^{1}\rightarrow S^{1},$ where $S^{1}=[0,1]$ is mapped into \emph{const}CS(%
\textbf{A}(1)). Here \emph{const} depends upon the gauge group being used so
that the combination \emph{const}CS(\textbf{A}(1)) is an integer multiplied
by $2\pi $ (e.g. see eq.(3.4)).

By looking at eq.(3.7) we notice that on the l.h.s. the integration is done
over\ the volume and time while on the r.h.s. only over 3-volume. To resolve
this difficulty, Donaldson suggests to use the following parametrization 
\begin{equation}
\mathbf{A(}t\mathbf{,y)=}tA(\mathbf{y})  \tag{3.8}
\end{equation}%
It should be noted that:

a) since $t\in $ [0,1] , we can parametrize $t$ as $t=f(x)$ , $-\infty
<x<\infty $ \ so that $\mathbf{A}(-\infty )\in a$ and $\mathbf{A}(+\infty
)\in b$. This is 3 dimensional interpretation of 3+1 dimensional instanton
construction.

b) in 4 dimensional interpretation, \ we have to keep in mind that in
temporal gauge $t$ plays a role of parameter. \ The well known (in physics
literature) 1-instanton solution (Polyakov 1987, Donaldson 2002)%
\begin{equation}
\left\vert F\right\vert =\frac{1}{\left( 1+r^{2}\right) ^{2}}  \tag{3.9}
\end{equation}%
obtained on \textbf{R}$^{4},$ when viewed in the 3+1 setting in which 
\textbf{R}$^{4}$ is replaced by $S^{3}\times \mathbf{R,}$ acquires different
(but equivalent) look 
\begin{equation}
\left\vert F\right\vert =\frac{4}{\cosh ^{2}(t)}.  \tag{3.10}
\end{equation}%
Now $r=r(f(t))$, $r^{2}(t)=x(t)^{2}+y(t)^{2}+z(t)^{2}+t^{2}.$ At $t=-\infty $
\ a path \textbf{r}$(t)$ begins at trivial connection for which we have $%
\left\vert F\right\vert =0.$ Then, for $t\rightarrow \infty $ the path winds
once (and only once!) in the space of connections $\mathcal{B}$ and returns
again to the flat connection.

\bigskip Eq.(3.8) allows us to replace the inequality (2.35) by 
\begin{equation}
\tint\limits_{Y}dv\mathbf{B}^{2}\geq \mathfrak{N}CS(\mathbf{\hat{A}}). 
\tag{3.11}
\end{equation}%
where $\mathfrak{N}$ is some constant. Since the connection $\mathbf{A}%
\equiv \mathbf{\hat{A}}$ must satisfy zero curvature eq.(2.18), we can
exploit this fact in order to write 
\begin{equation}
d\mathbf{\hat{A}=-\hat{A}\wedge \hat{A}.}  \tag{3.12}
\end{equation}%
Using this result in the CS functional we obtain (up to a constant) 
\begin{equation}
CS(\mathbf{\hat{A}})=-\frac{1}{3}\tint\limits_{Y}tr(\mathbf{\hat{A}}\wedge 
\mathbf{\hat{A}}\wedge \mathbf{\hat{A}).}  \tag{3.13}
\end{equation}%
Such a form of the CS functional is used in physics literature (Manton and
Sutcliffe 2007). There is yet another way to write the same functional.
Indeed, using eq.(3.12) in eq.(2.16) we obtain as well%
\begin{equation}
CS(\mathbf{\hat{A}})=\frac{1}{3}tr\tint\limits_{Y}\mathbf{\hat{A}}\wedge d%
\mathbf{\hat{A}}.  \tag{3.14}
\end{equation}%
By combining eq.s(3.11) and (3.14) we obtain the result of central
importance for our work:%
\begin{equation}
\tint\limits_{Y}dv\mathbf{B}^{2}\geq \mathbb{N}\tint\limits_{Y}tr(\mathbf{%
\hat{A}}\wedge d\mathbf{\hat{A}}),  \tag{3.15}
\end{equation}%
where $\mathbb{N}$ is yet another positive constant. \ The discussion of
this inequality starts in the next section.\bigskip

\textbf{4. Arnol'd inequality, Ranada's electromagnetic }

\ \ \ \ \textbf{tensor and contact geometry of monopole\bigskip \bigskip s
and dyons}

\textit{4.1. Arnol'd inequality \bigskip }

Using results of appendix C, the Abelian version of inequality (3.15) can be
rewritten in hydrodynamics language (Kholodenko 2013) where it is known as
Arnol'd inequality (Arnol'd and Khesin 1998) 
\begin{equation}
\tint\limits_{Y}dv\mathbf{v}^{2}\geq \mathbb{N}\tint\limits_{Y}dv\mathbf{%
v\cdot }\left( \mathbf{\nabla \times v}\right) .  \tag{4.1a}
\end{equation}%
This inequality can be equivalently written as 
\begin{equation}
\mathcal{E}[\mathbf{v}]=\tint\limits_{Y}dv\mathbf{v}^{2}\geq \mathbb{C}%
\tint\limits_{Y}dv(\mathbf{v\cdot }curl^{-1}\mathbf{v)\equiv }\mathbb{C}%
\mathcal{H}[\mathbf{v}]\mathbf{.}  \tag{4.1b}
\end{equation}%
The constant $\mathbb{C}$ is \ determined in Arnol'd and Khesin's book. It
will be also discussed further below. Thus, the helicity $\mathcal{H}[%
\mathbf{v}]$ provides the lover bound for the (kinetic) energy functional $%
\mathcal{E}[\mathbf{v}].$ Independently, one can pose the problem: Find the
minimum of $\mathcal{E}[\mathbf{v}]$ and the extremals among the velocity
field obtained from a given divergence-free field $\mathbf{v}$ by the action
of volume-preserving diffeomorphisms. \ Arnol'd proved the following\bigskip

\textbf{Theorem 4.1.} \textit{The extremals of the just stated problem are
divergence-free vector fields that commute with their vorticities. In
particular, they coincide with steady Euler flow in }$Y$\textit{.\bigskip }

The content of this theorem should be understood as follows. Define the
commutator (the analog of Poisson brackets) for \ the vector fields $\varphi 
$ and $\psi $ as $\{\mathbf{\varphi },\mathbf{\psi }\}=(\mathbf{\varphi
\cdot \nabla )\psi -}(\mathbf{\psi \cdot \nabla )}\mathbf{\varphi }$ then,
it can be demonstrated that $\{\mathbf{\varphi },\mathbf{\psi }\}=curl(%
\mathbf{\varphi }\times \mathbf{\psi )-\varphi (}$div$\mathbf{\psi )+\psi (}$%
div$\mathbf{\varphi }\mathbf{).}$ By applying this identity to the
divergence-free fields we obtain, 
\begin{equation*}
\{\mathbf{\varphi },\mathbf{\psi }\}=curl(\mathbf{\varphi }\times \mathbf{%
\psi ).}
\end{equation*}%
If \ the fields $\mathbf{\varphi }$ and $\mathbf{\psi }$ commute, we get $%
\mathbf{\varphi }\times \mathbf{\psi =\nabla }\alpha \mathbf{.}$ When \ this
is rewritten in hydrodynamic language, this result acquires the form of
equation describing stationary Euler flow 
\begin{equation}
\mathbf{v}\times curl\mathbf{v}=\nabla \alpha  \tag{4.2}
\end{equation}%
with the Bernoulli function $\alpha =\mathcal{P}\mathbf{+}\frac{\mathbf{v}%
^{2}}{2}$. This result is in accord with eq.(C.3a). When translated into
electromagnetic language, the analogous result is given by eq.(C.7) of
Appendix C 
\begin{equation}
\mathbf{E+v\times B=-\nabla }\Phi  \tag{C.7}
\end{equation}%
But, since thus far we were working with the source-free Maxwell fields the
particle(or fluid) velocity \textbf{v }in\textbf{\ }(C.7) creates some
problem in establishing fluid mechanics-electrodynamics correspondence. This
issue was studied first by Newcomb (1958). The latest attempt at resolving
this issue can be found in the paper by van Enk (2013).

Before switching to electrodynamics, it is educational to discuss the
obtained results in hydrodynamics setting. This is justified because in this
case the solutions to eq.(4.2) were analyzed rigorously by Arnol'd (Arnol'd
and Khesin 1998).

The analysis begins with the case $\alpha =const$ in eq.(4.2). \ This
condition leads to the so called \textsl{force-free} \ family of solutions 
\footnote{%
Note the difference in terminology. The "force-free" termin is used in
magnetohydrodynamics while in hydrodynamics the termin "Beltrami" is used
instead.}. These are the solutions of the eigenvalue problem 
\begin{equation}
curl\mathbf{v=}\kappa \mathbf{v}\text{ }  \tag{4.3}
\end{equation}%
where the scalar function $\kappa $ may or may not be a constant. Suppose it
is not a constant, then using eq.(4.3) we obtain: \ div$\kappa \mathbf{%
v=v\cdot \nabla }\kappa \mathbf{=}0.$ We would like to remind to our readers
what this equation actually means. If $\kappa =\kappa (x,y,z)=const$ is an
equation describing surface and if \textbf{r}(t) is some trajectory \textbf{r%
}(t)=\{x(t),y(t),z(t)\} on this surface, then $\frac{d}{dt}\kappa
(x(t),y(t),z(t))=v_{x}\kappa _{x}+v_{y}\kappa _{y}+v_{z}\kappa _{z}=\mathbf{%
v\cdot \nabla }\kappa \mathbf{=}0.$ That is the condition $\mathbf{v\cdot
\nabla }\kappa \mathbf{=}0$ means that the fluid velocity \textbf{v} is
always tangential to the surface $\kappa (x,y,z)=const$ , e.g. read
(Dubrovin et al 1984).\ Since the vector field \textbf{v} is assumed to be
nowhere vanishing, the surface $\kappa (x,y,z)=const$ can only be torus T$%
^{2}$. The field lines of \textbf{v} on T$^{2}$ should be closed if $const$
is rational number. Thus, the force-free condition, eq.(4.3), provides us
with the condition for existence of all possible torus knots for rational $%
\kappa ^{\prime }s$ (Gilbert and Porter 1994).\bigskip

\textbf{Corollary 4.2.} \textit{The force-free vector fields defined by
eq.(4.3) minimize} $\mathcal{E}[\mathbf{v}].\bigskip $

Indeed, using eq.(4.3) in (4.1a) produces 
\begin{equation*}
\tint\limits_{Y}dv\mathbf{v}^{2}\geq \kappa \mathbb{N}\tint\limits_{Y}dv%
\mathbf{v}^{2}
\end{equation*}

so that the equality is achieved when $\kappa \mathbb{N=}1$. Details of the
proof are given in

(Arnol'd and \ Khesin 1998).\bigskip

Being armed with hydrodynamical results, now we must look at the analogous
interpretation for the elelectrodynamical eq.(C.7). \ \ Evidently, the
electrodynamics problem is reducible to that we just discussed if \ we
select the potential $\Phi $ in such a way that $\mathbf{\nabla }\Phi +%
\mathbf{E}=0.$ This choice happens to be permissible as explained in
Appendix D. If this is so, we have to decide what to do with the equation $%
\mathbf{v\times B=0}$. Fortunately, this equation was studied in (Kholodenko
2013) so that here we provide only the summary of results.

First, we notice that $\mathbf{B=\nabla \times A,}$ $\nabla \cdot \mathbf{A}%
=0.$\ Second, we need to assume that $\mathbf{v}$=$\pm \gamma \mathbf{A}$
where $\gamma $ is some real constant. Using these facts, we obtain: \ 
\begin{equation}
\mathbf{A}\times \mathbf{\nabla \times A=0.}  \tag{4.4}
\end{equation}%
This leads us back to the equation analogous to eq.(4.3), that is we have 
\begin{equation}
\mathbf{\nabla \times A=}\kappa \mathbf{A.}  \tag{4.5a}
\end{equation}%
Clearly, now we can apply all results related to eq.(4.3) to the present
case. But, in addition, by applying the curl operator to both sides we
obtain as well 
\begin{equation}
\mathbf{\nabla }\times \mathbf{B=}\kappa \mathbf{B}  \tag{4.5b}
\end{equation}%
which is the force-free equation. Eq.(4.5a) can be altrernatively rewritten
in the form of the London-type equation of superconductivity\footnote{%
For the proof that this is indeed the London-type equation, please, read
(Kholodenko 2013), page 3.} \ 
\begin{equation}
\mathbf{\nabla \times v=}\kappa \mathbf{v}  \tag{4.5c}
\end{equation}%
and, if this is so, the vortices in superconductors and superfluids fall in
the same category as beams of light. Thus, the already discussed connection
beween monopoles and superconducting rings comes to play immediately.\medskip

\textit{4.2. Helicity and contact geometry\bigskip }

Obtained results bring us into position \ when we can discuss Ranada's
electromagnetic tensor. The most optimal way to reobtain Ranada's tensor is
trough use of the helicity. Since Ranada uses Clebsch variables, we begin
with the observation made by Bretheron (1970). Written in electromagnetic
language, his argument goes as follows (appendix D). Let $\mathbf{A}=\nabla
\varphi +\alpha \nabla \beta ,$ so that $\mathbf{B}=\nabla \times \mathbf{%
A=\nabla }\alpha \mathbf{\times \nabla }\beta .$ Using these facts, the
helicity reads as 
\begin{eqnarray}
\mathcal{H}[\mathbf{A}] &=&\dint\limits_{V}dv\mathbf{A}\cdot \mathbf{B=}%
\dint\limits_{V}dv\{\nabla \varphi \cdot \mathbf{\nabla }\alpha \mathbf{%
\times \nabla }\beta \}  \notag \\
&=&\dint\limits_{V}dv\nabla \cdot \{\varphi \mathbf{\nabla }\alpha \mathbf{%
\times \nabla }\beta \}  \notag \\
&=&\dint\limits_{\Sigma }d\vec{\Sigma}\cdot (\mathbf{\nabla }\alpha \mathbf{%
\times \nabla }\beta )\varphi =0.  \TCItag{4.6}
\end{eqnarray}%
The last line comes from identifications: $\mathbf{A}\rightleftarrows 
\mathbf{v},\mathbf{B}\rightleftarrows \mathbf{\vec{\omega}}$ and observation
that $\mathbf{\vec{\omega}\cdot \vec{\Sigma}=0}$ on the surface of the
vortex tube (Saffman 1995). Thus, it looks like use of Clebsch variables is
restricted to non entangled flows. It can be also demonstrated that Clebsch
variables cannot be used when the vorticity (or magnetic field) vanishes in
some regions of space (Boozer 2010), (Graham and Heney 2000). The way out of
this difficulty was found already in the classical paper by Seliger and
Whitham (1968). The same result was rediscovered by Goncharov and Pavlov
(1997) and later by Yoshida (2009). Alternative treatment was given by
Kuznetsov and Mikhailov (1980).

For the sake of space, we shall discuss only results by Yoshida (2009) since
they can be immediately linked with contact geometry. \ Yoshida
notices/proves that for an arbitrary vector field \textbf{u} it may not be
possible to find 3 scalars (scalar functions) $\alpha ,\beta ,\varphi $ so
that representation $\mathbf{u}=\mathbf{\nabla }\varphi +\alpha \mathbf{%
\nabla }\beta $ is defined \textit{globally} in space. That is the map $%
(u_{1},u_{2},u_{3})\rightarrow (\varphi ,\alpha ,\beta )$ may not be
injective. The situation can be corrected if instead \ the generalized form $%
\mathbf{u}=\nabla \varphi +\dsum\nolimits_{J=1}^{\delta }\alpha _{J}\nabla
\beta _{J}$ is being used with $\delta =D-1$ and $D$ being the
dimensionality of space. Incidentally, when $D=3$, we obtain 
\begin{equation}
\mathbf{u}=\nabla \varphi +\alpha _{1}\nabla \beta _{1}+\alpha _{2}\nabla
\beta _{2}  \tag{4.7}
\end{equation}%
This result was obtained for the first time by Seliger and Whitham (1968),
and rediscovered later by Goncharov and Pavlov (1997). \ Following Yoshida
(2009) and also Marsden and Weinstein (1983) we rewrite eq.(4.7) in terms of
differential 1-form 
\begin{equation}
du=d\varphi +\dsum\nolimits_{j=1}^{m}\alpha _{j}d\beta _{j}  \tag{4.8}
\end{equation}%
easily recognizable as contact 1-form (Kholodenko 2013, Geiges 2008). For
readers with standard physics equation we notice that the familiar from
mechanics relation $p_{i}$=$\frac{\partial S}{\partial q_{i}}$ can be
rewritten as the kernel (that is $\omega =0)$ of the contact 1-form $\omega
=dS+\dsum\nolimits_{j=1}^{m}p_{j}dq_{j}.$ Because of this, it is clear that $%
\omega $ should be invariant with respect to canonical transformations as
stated in appendix D. But, in addition, \ in contact geometry there are
contactomorphic transformations which can be understood as follows. Let $m=1$
in the above 1-form $\omega $ and consider a transformation $%
(q,S,p)\rightarrow (q^{\prime },S^{\prime },p^{\prime }).$ It is considered
as \textit{contactomorphic} \ if there is a function $\rho :\mathbf{R}%
^{3}\rightarrow \mathbf{R}$ which is nowhere zero and such that%
\begin{equation}
dS^{\prime }+p^{\prime }dx^{\prime }=\rho (q,S,p)(dS+pdx).  \tag{4.9}
\end{equation}%
The connection with symplectic mechanics is achieved via relation 
\begin{equation}
d\omega =\dsum\nolimits_{j=1}^{m}dp_{j}\wedge dq_{j}  \tag{4.10}
\end{equation}%
It is clear that the smallest dimension of the symplectic manifold is 2.
Accordingly, \ the smallest dimension of the contact manifold is 3. \ A 
\textit{contact structure} on 3-manifold $M$ \ is a smoothly varying plane
field $\xi $ (made of planes $\omega =0)$ which is completely nonintegrable%
\footnote{%
For details regarding this concept, please, consult (Kholodenko 2013, Geiges
2008)}, that is 
\begin{equation}
\omega \wedge d\omega \neq 0.  \tag{4.11}
\end{equation}%
Consider now the simplest Clebsch representation written in 1-form language,
e.g. $\omega =d\varphi +\alpha d\beta $. Then $d\omega =d\alpha \wedge
d\beta $ is the symplectic volume (area) 2-form. It can be defined both in 
\textbf{R}$^{2}$ and on $S^{2}.$ $\ $Evidently$,S^{2}$ is just one point
compactification of \textbf{R}$^{2}.$ Below we shall investigate both cases%
\footnote{%
Albeit not with equal amount of details}. For now, we notice that the
Bianchi identity $d\mathbf{F}=0$ for Maxwell's eq(2.8a) can be interpreted
symplectically. Indeed, if we label the 2-form in eq.(4.10) as \textbf{F},
then, surely, for such defined \textbf{F }we obtain\textbf{\ }$d\mathbf{F}%
=0. $ But we already know that $d\omega =d\alpha \wedge d\beta $. Therefore
we reobtain back $\mathbf{\nabla }\alpha \mathbf{\times \nabla }\beta =%
\mathbf{B}$ as required. Notice that it is this \textbf{B} which is present
in the formula for helicity, eq.(4.6). \ Now we can rewrite the formula for
helicity $\mathcal{H}[\mathbf{A}]$ \ alternatively as 
\begin{equation}
\mathcal{H}[\mathbf{A}]=\dint\limits_{V}\omega \wedge d\omega .  \tag{4.12a}
\end{equation}%
in accord with eq.(3.15). By doing so, we run into contradiction: on one
hand, in view of eq.(4.11) $\mathcal{H}[\mathbf{A}]$ must be nonzero, while
on another, according to eq.(4.6) it should be zero. The best way out is by
relating this problem with that for Dirac's monopole.\bigskip

\textit{4.3. \ From Dirac monopoles to Ranada's dyons\bigskip }

In our discussion of this topic we follow (Kholodenko 2013, Arnol'd and
Khesin 1998 and Ryder 1980). We begin with Example 1.19 in (Arnol'd and
Khesin 1998). Let $\mathbf{F}$ be arbitrary area 2-form on $S^{2}$
normalized by the condition $\dint\nolimits_{S^{2}}\mathbf{F}=1.$ Such a
form is closed on $S^{2}$ but it is not exact as we shall demonstrate
momentarily. Consider a (Hopf) mapping $\pi :S^{3}\rightarrow S^{2}$, then
the pullback $\pi ^{\ast }\mathbf{F}$ produces 2-form which is exact on $%
S^{3}$. That is if\ there is 1-form $\omega $ on $S^{3}$ such that $d\omega
=\pi ^{\ast }\mathbf{F},$then 
\begin{equation}
\mathcal{H}[\mathbf{A}]=\dint\limits_{V}\omega \wedge \pi ^{\ast }\mathbf{F}
\tag{4.12b}
\end{equation}%
is the Hopf invariant. Since it is a winding (linking) number it is an
integer. This statement is \ formulated as Proposition 1.20. in the book by
Arnol'd and Khesin (1998). No proof of this proposition is given in this
reference. Only a hint. A proof is given in (Kholodenko 2013). Special cases
are \ discussed below.

The 2-form on $S^{2}$ is given in (Ryder 1980) in the context of Dirac
monopole. When normalized in accord with Arnol'd and Khesin, this form is
given by $\mathbf{F}=\frac{1}{4\pi }\sin \theta d\theta \wedge d\phi .$ As
it is shown in (Eguchi et al 1980) an attempt to represent $\mathbf{F}$ as $%
\mathbf{F}=dA$ on the entire 2-sphere is useless. Instead it is possible to
subdivide $S^{2}$ into the Nothern and Southern parts so that $\mathbf{F}%
=dA_{\pm }$ for these parts respectively with 
\begin{equation}
A_{\pm }=\frac{1}{4\pi r}\frac{1}{z\pm r}(xdy-ydx).  \tag{4.13a}
\end{equation}%
In spherical coordinates and for the sphere of unit radius we obtain instead 
\begin{equation}
A_{\pm }=\frac{1}{4\pi }(\pm 1-\cos \theta )d\phi .  \tag{4.13b}
\end{equation}%
Both expressions are easily recognizable as vector potentials associated
with the Dirac monopole. Using these potentials we obtain indeed $\mathbf{F}%
=dA_{\pm }.$

\ The above results were obtained for $S^{2}.$ Now we would like to
demonstrate that for \textbf{R}$^{2}$ the results are different. Indeed,
take 1-form $\omega =dS+pdx.$ From here $\mathbf{F}=d(pdx)=dp\wedge dx.$
Accordingly, $\dint\nolimits_{R^{2}}dp\wedge dx=\dint\nolimits_{\partial
R^{2}}pdx=0$ since any contour can be squeezed to zero (Cauchy's theorem).
Alternatively, this result is coming from the fact that the relationship $%
p_{i}$=$\frac{\partial S}{\partial q_{i}}\equiv f(x_{1},...,x_{i},...)$ for $%
p_{i}$ being assigned describes the \textit{Lagrangian} manifold. Therefore,
for such manifold we must have $\dsum\nolimits_{i}\doint p_{i}dx_{i}=0$
since $dp\wedge dx=\dfrac{dp}{dx}dx\wedge dx.$\ Thus, \textbf{R}$^{2}$ is
not \ good for our purposes and we have to work with $S^{2}.$ Next we have
to find 1-form $\omega $ on $S^{3}$ such that $d\omega =\pi ^{\ast }\mathbf{F%
}$ where $\mathbf{F}=\frac{1}{4\pi }\sin \theta d\theta \wedge d\phi \mathbf{%
.}$ The 1-form $\omega $ on $S^{3}$ is contact, as is well known. Therefore,
(up to a constant) its standard form is given by%
\begin{equation}
\omega =\dsum\nolimits_{i=1}^{2}(x_{i}dy_{i}-y_{i}dx_{i}).  \tag{4.14}
\end{equation}%
Using it, we obtain (again, up to a constant) 
\begin{equation}
d\omega =2(dx_{1}\wedge dy_{1}+dx_{2}\wedge dy_{2}).  \tag{4.15a}
\end{equation}%
Next, we introduce the complex variables $z_{0}=x_{1}+iy_{1}$ and $%
z_{1}=x_{2}+iy_{2}$ . In terms of these variables the 2-form $d\omega $ is
written now (up to a constant) as (Bott and Tu 1982) 
\begin{equation}
d\omega =\frac{i}{2\pi }(z_{1}dz_{0}-z_{0}dz_{1})\wedge (\bar{z}_{1}d\bar{z}%
_{0}-\bar{z}_{0}d\bar{z}_{1})  \tag{4.15b}
\end{equation}%
By keeping in mind that the equation for the 3-sphere of unit radius is 
\begin{equation*}
\left\vert z_{0}\right\vert ^{2}+\left\vert z_{1}\right\vert ^{2}=1
\end{equation*}%
we can rewrite eq.(4.15b) as 
\begin{equation}
d\omega =\frac{i}{2\pi }\frac{(z_{1}dz_{0}-z_{0}dz_{1})\wedge (\bar{z}_{1}d%
\bar{z}_{0}-\bar{z}_{0}d\bar{z}_{1})}{\left( \left\vert z_{0}\right\vert
^{2}+\left\vert z_{1}\right\vert ^{2}\right) ^{2}}  \tag{4.15c}
\end{equation}%
In such a form it is valid on $S^{3}$. Now it remains to demonstrate that
this result can be presented as $d\omega =\pi ^{\ast }\mathbf{F.}$ For this
purpose, following (Bott and Tu 1982) it is sufficient to define $%
z=z_{0}/z_{1}\in S^{2}$ so that in terms of $z$ the above 2-form can be
rewritten as 
\begin{equation}
\mathbf{F}=\frac{i}{2\pi }\frac{dz\wedge d\bar{z}}{\left( 1+\left\vert
z\right\vert ^{2}\right) ^{2}}.  \tag{4.16}
\end{equation}%
This \ is the 2-form suggested by Ranada (1989, 1992) for the
electromagnetic tensor. The connection with electromagnetism follows from
the fact that by design this 2-form satisfy the Bianchi identity $d\mathbf{F}%
=0$ revealing its symplectic origin as explained in subsection 4.2.

To complete this subsection we still have to discuss several issues. We
begin with demonstration that the 2-form $\mathbf{F}=\frac{1}{4\pi }\sin
\theta d\theta \wedge d\phi $ and that given in eq.(4.16) are indeed
equivalent. For this purpose, we notice that in spherical coordinates $%
z=cos\theta $ so that (up to sign) previous result can be rewritten as $%
\mathbf{F}=\frac{1}{4\pi }d\phi \wedge dz$ . This is the standard symplectic
2-form on $S^{2}$ discussed in detail in (Kholodenko 2013). \ By design
(e.g. read above) it obeys the normalization condition 
\begin{equation*}
\dint\nolimits_{S^{2}}\mathbf{F=}\frac{1}{4\pi }\dint\limits_{0}^{2\pi
}d\phi \dint\limits_{0}^{2}dz=1.
\end{equation*}%
The 2-form in eq.(4.16) is also discussed in detail in (Kholodenko 2013). It
is the \ standard Fubini-Study 2-form on $S^{2}$ \ \ In this case the
2-sphere as symplectic manifold is being interpreted as complex projective
line \textbf{CP}$^{1}($recall that $z=z_{0}/z_{1})$ (Arnol'd 1989). In polar
coordinates it can be rewritten as (up to normalization constant) 
\begin{equation}
\mathbf{F}=\frac{1}{2\pi }\frac{d\phi \wedge rdr}{\left( 1+r^{2}\right) ^{2}}
\tag{4.17}
\end{equation}%
It coincides with the 2-form $\Omega $ presented in the book by Arnol'd
(1989) as a Problem 1 on page 347. Evidently the 2-form defined in
eq.s(4.16),(4.17) and $\mathbf{F}=\frac{1}{4\pi }d\phi \wedge dz$ are
equivalent. This is so because symplectic mechanics can be rewritten in
terms of dynamics in complex projective space \textbf{CP}$^{n}$ (Arnol'd
1989). The sphere $S^{2}$ is just \textbf{CP}$^{1}.$ These projective spaces
are examples of complex \ K\"{a}hler manifolds. From here, every K\"{a}hler
manifold is also a symplectic manifold. The opposite is not true as it was
demonstrated by Thurston (1976). In the present case the 2-form, eq.(4.16),
is nesessarily symplectic and, in view of the Darboux theorem, it can be
brought to the standard symplectic 2-form, that is to $\mathbf{F}=\frac{1}{%
4\pi }d\phi \wedge dz.$ To demonstrate this explicitly, we have to normalize
the form in eq.(4.17) correctly. That is we have to make sure that $%
\dint\nolimits_{S^{2}}\mathbf{F}=1.$This leads us to determination of the
constant $C$ using the equation 
\begin{equation}
\frac{C}{2\pi }\dint\limits_{0}^{2\pi }d\phi \dint\limits_{0}^{\infty }\frac{%
rdr}{\left( 1+r^{2}\right) ^{2}}=1.  \tag{4.18}
\end{equation}%
Evidently,%
\begin{equation*}
\dint\limits_{0}^{\infty }\frac{rdr}{\left( 1+r^{2}\right) ^{2}}=\frac{1}{2}%
\dint\limits_{1}^{\infty }\frac{du}{u^{2}}=\frac{1}{2}
\end{equation*}%
\qquad Therefore, $C=2$. Using this result, we can now compare these
2-forms. Specifically, we obtain:%
\begin{equation*}
\frac{dz}{2}=\frac{du}{u^{2}}\text{ or }\frac{1}{2}z=-\frac{1}{u}+g
\end{equation*}%
with $g$ being yet another constant. Since $u=1+r^{2}$, and for the sphere
of unit radius $z$ ranges between $0$ and $2$, the constant $g$ \ is easily
determined. Specifically, $g=1.$ Thus these 2-forms are indeed equivalent.
In view of this fact, the connection of \ Randada's results with those for
Dirac monopoles (Ryder 1980) is established. \ It will be reobtained at more
advanced level in part II. Since \ Ranada (1992) uses \ two types of\
helicities- electric and magnetic- all results for magnetic (Dirac)
monopoles can be transferred without change to the electric-type monopoles.
If the electric monopole is located at the same place of spacetime as the
magnetic monopole, we obtain a dyon. That is \ a hypotetical particle
possessing simultaneously electric and magnetic charges. \ Thus, in accord
with results of appendix A, the electromagnetic field designed by Ranada
(1989) is that \ originating from dyons. As it was explained in section
2.2., both monopoles and dyons are made of interlocked magnetic and electric
Hopf-type rings. Thus, in accord with Ranada, there is no need to use the
actual charges for description of monopoles or dyons. Furthermore, \ in part
II we shall be dealing with complements of knots/links. According to
Thurston, now proved, geometrization conjecture there are 8 geometries for
3-manifolds. As in two dimensional case, one first have to study spaces of
positive, zero and negative curvatures and then groups of isometries in
these spaces. In 3 dimensions situation changes since complements of
knots/links \ initially embedded in $S^{3}$ create spaces of positive, zero
and negative curvatures depending on knot link topology, that is on the
fundamental group of the knot/link complement (Scott 1983). Thus, if we
believe Einstein, only physical masses can create curvatures. Now, with
Dirac monopoles interpreted as interlocked magnetic rings and Ranada's dyons
need in masses \ not only disappear but even becomes problematic.

Next, \ we need to demonstrate that the 2-forms given by eq.s(4.7),(4.8) and
eq.(4.14) are equivalent. This is so because in the 1st case we have the
1-form $\omega =d\varphi +\alpha _{1}d\beta _{1}+\alpha _{2}d\beta _{2}$ so
that $d\omega $ (up to a constant) coincides with that given in eq.(4.15a).
This means that using \ such 1-form we can recover Ranada's \textbf{F . \ }%
Just defined\ contact 1-form lives in \textbf{R}$^{5}$ while that given by
eq.(4.14) lives on $S^{3}.$ But $S^{3}$ is embedded in \textbf{R}$^{4}$ and 
\textbf{R}$^{4}$ is symplectic space with symplectic 2-form given by
eq.(4.15a). It is standard result of contact geometry (Geiges 2008,
Kholodenko 2013) that it is always possible to represent the contact 1-form $%
\alpha $ living in \textbf{R}$^{5}$ as 
\begin{equation}
\alpha =dz+\dsum\nolimits_{i=1}^{2}(x_{i}dy_{i}-y_{i}dx_{i})  \tag{4.19a}
\end{equation}%
More accurately, it can be demonstrated (Geiges 2008), page 52, that $\alpha 
$ is contactomorphic to%
\begin{equation}
\tilde{\alpha}=dz+\dsum\nolimits_{i=1}^{2}x_{j}dy_{j}  \tag{4.19b}
\end{equation}%
From here it follows that 1-forms living in \textbf{R}$^{5}$ and on $S^{3}$
are fully compatible (contactomorphic). This result provides instant proof
of results of Seliger and Whitham (1968), Goncharov and Pavlov (1997) and by
Yoshida (2009).\medskip \medskip

\textbf{5. Discussion \bigskip }

We begin by noticing that Ranada's 2-form, eq.(4.16), is identical with the
2-form, eq.(3.9), obtained via instanton method. This observation is in
formal accord with results by Trautman (1977) \ who obtained this result
only implicitly by using different arguments. From Section 3 we know that
such one-instanton solution should have instanton (winding) number equal to
one. This is surely the case for the helicity $\mathcal{H}[\mathbf{A}],$ eq$%
.(4.12a),$ if it is computed using the 1 and 2-forms described in the
previous \ section. Computations done by Bott and Tu (1982), by Ryder (1980)
and, by Kuznetsov and Mikhailov (1980), confirm this. \ Questions arise: a)
Can one construct multiinstanton solutions in the Abelian case if such
solutions do exist in the non-Abelian Y-M case (Manton and Sutcliffe 2007)?
b) What kinds of knots/links, if any, can be associated with the
multiinstanton-type solutions if such solutions exist? c) How method of
Floer discussed in Section 2 can accommodate the multiinstanton solutions?
d) In the case if the above questions are answered affirmatively, how one
then should relate them to the fact (Witten 1989) that the Abelian CS field
theory is known to be capable of describing only the Hopf-type links and
torus-type knots/links? \ We shall address some of these problems in the
companion publication, part II. \ In it we shall make a heavy use of the
force-free (or Beltrami) eq.(4.3). As results of the appendix D indicate,
this equation is both relativistically and gauge invariant which makes its
treatment apparently easier\footnote{%
In fact, not a bit easier as references to part II indicate.}. From the
discussions presented in this paper it should become clear that this
equation is inseparable from the Arnol'd inequality and, therefore from the
helicity. The helicity by design is the product of contact
geometry/topology. \ Since the existence of nonzero helicity is indicative
of existence of magnetic/electric monopoles, it follows that electric and
magnetic monopoles can be described purely geometrically. Accordingly, to
move beyond current results in physics literature describing knotted beams
of light will require us to go much deeper into formalism of contact
geometry/topology. For instance, in appendix D we discussed constructions
based on use of Clebch variables $\alpha $ and $\beta $ while in the section
4.2. we introduced eq.(4.7) which requires two copies of $\alpha ^{\prime }s$
and $\beta ^{\prime }s.$ If this requirement is ignored, we would end up
with negative result summarized in eq.(4.6). Furthermore, Lund and Regge
(1976) \ demonstrated that while eq.s(D.3a) and (D.3b) are correct from the
Galilean perspective, they are not reparametrization-invariant. When
reparamerization invariance is enforced, we end up with vortices modelled by
relativistic strings. These should be knotted/linked. Recall that the
connection between strings and vortices, had began with ground breaking work
by Nielsen and Olessen (1973). Faddeev and Niemi and many other researchers
studied such knotted/linked strings within the framework of abelianized QCD.
Details and many up-to date references on the Faddeev-Skyrme model
describing such knotted/linked structures can be found in (Kholodenko 2011
and Kholodenko 2013). In view of dyonic interpretation of Ranada's results
it is very appropriate to finish this part of our work with the following
observations.

Both in Y-M and Einsteinian gravity the problem of treatment of extended
bodies is extremely difficult, e.g. read \ (Kholodenko 2013), page 97 and
references therein. Einstein was always unhappy with the right hand side of
his equations since this side contained masses. The masses do not fit \ well
into his geometrical formalism. It is well known that both Y-M and gravity
have monopole-like solutions. Both these theories can be described
topologically in terms of loops, knots/links. \ Depending upon knot/link
topology (that is on the first fundamental group of \ the knot/link
complement \ in S$^{3})$ these complementary spaces will have positive, zero
or negative curvatures. According to Einstein curvatures can only be created
by masses. In this paper, part I, and in its companion, part II, we provide
enough evidence that masses are not nesessary. Interesting attempt to
describe the standard model in terms of knots/links \ is proposed in the
series of papers by Finkelstein, e.g. see (Finkelstein 2005). \ Alternative
treatment of both gravity and the standard model in terms of knots and links
is given in (Kholodenko 2011).\bigskip

\bigskip

\textbf{Acknowledgement }\ The author would like to thank Dr. Zurab K.
Silagadze (Budker Institute, Novosibirsk, Russia) for many \ discussions
regarding the content of this paper and for some technical
assistance.\bigskip

\textbf{Appendices\bigskip }

\textit{Appendix A.}\textbf{\ }\textit{Pre quantization of Y-M fields and
electric-magnetic duality\bigskip }

Without account of gauge constraints, the action functional $S$ for both
Abelian and non-Abelian Y-M fields in space-times of Minkowski signature is
given by (Frankel 1997) 
\begin{equation}
S=\frac{1}{2}\tint\limits_{-\infty }^{\infty }dt\tint\limits_{Y}dv(3)[%
\mathbf{E}^{2}-\mathbf{B}^{2}]  \tag{A.1}
\end{equation}%
Naturally, it coincides with the well known result for the Maxwell fields
(Landau and Lifshitz 1975) \ For these fields $\mathbf{B}=$ $\mathbf{\nabla }%
\times \mathbf{A}$ and $\mathbf{E}=-\frac{\partial }{\partial t}\mathbf{A}-%
\mathbf{\nabla }\varphi ,$ as is well known, $A_{0}=\varphi .$ For both the
Abelian and non-Abelian Y-M fields it is convenient to use the temporal
gauge in which $A_{0}=0.$ In this gauge the above action can be rewritten in
the form%
\begin{equation}
S[\mathbf{A}]=\frac{1}{2}\tint\limits_{-\infty }^{\infty
}dt\tint\limits_{Y}dv(3)[\mathbf{\dot{A}}^{2}-\left( \mathbf{\nabla }\times 
\mathbf{A}\right) ^{2}],  \tag{A.2}
\end{equation}%
where $\mathbf{\dot{A}=}\frac{\partial }{\partial t}\mathbf{A.}$ From the
condition $\frac{\delta S[\mathbf{A}]}{\delta \mathbf{A}}=0$ it follows
that: $\frac{\partial }{\partial t}\mathbf{E=\nabla }\times \mathbf{B.}$ The
definition of $\mathbf{B}$ guarantees the validity of the condition $\mathbf{%
\nabla }\cdot \mathbf{B}=0.$ At the same time, from the definition of $%
\mathbf{E}$ we obtain another Maxwell's equation: $\frac{\partial }{\partial
t}\mathbf{B=-\nabla }\times \mathbf{E.}$ The question remains, nevertheless.
Will these results reproduce the remaining Maxwell's equation $\mathbf{%
\nabla }\cdot \mathbf{E}=0?$ This equation is essential for correct
formulation of the Cauchy problem for these fields\footnote{%
Essentials on the Cauchy problem for Maxwell and gravity fields can be found
in our work (Kholodenko 2011), version 1.}. If it is satisfied for $t=0$, it
will be also satisfied for $t>0$. Analysis shows, however, that for $t=0$
the existence of this equation is \textit{not} \ a consequence of the
remaining equations. Accordingly, it should be imposed as an independent
condition. This is primary source of some technical difficulties. Without
describing them in full, we sketch some key steps. Specifically, let the
vector field $\mathbf{A}$ be decomposed as $\mathbf{A}=\mathbf{A}_{\parallel
}+\mathbf{A}_{\perp }$. Since $\mathbf{E}=-\frac{\partial }{\partial t}%
\mathbf{A,}$ we obtain as well $\mathbf{\nabla }\cdot (\mathbf{E}_{\parallel
}\mathbf{+E}_{\perp }\mathbf{)}=0.$ By design, 
\begin{equation}
\mathbf{\nabla }\cdot \mathbf{E}_{\perp }=0  \tag{A.3}
\end{equation}%
At the same time $\mathbf{\nabla }\cdot \mathbf{E}_{\parallel }$ remains to
be defined by the initial and boundary data. In view of these conventions it
is always possible to choose $\mathbf{A}_{\parallel }=0$ and to use only $%
\mathbf{A}_{\perp }$ for quantization. Thus, $S[\mathbf{A}]$ defined in
eq.(A.2) can be finally rewritten as 
\begin{equation}
S[\mathbf{A}_{\perp }]=\frac{1}{2}\tint\limits_{-\infty }^{\infty
}dt\tint\limits_{Y}dv(3)[\mathbf{\dot{A}}_{\perp }^{2}-\left( \mathbf{\nabla 
}\times \mathbf{A}_{\perp }\right) ^{2}].  \tag{A.4}
\end{equation}%
so that in eq.(2.10b) we have in fact $dv(3,1)=dv(3)dt.$ In such a form it
is being used as an action functional in the exponent of the path integral
for the Y-M fields in both Abelian and non-Abelian settings (Huang 1982),
page 152, (Deser and Teitelboim 1976, Donaldson 2002), We shall not write
the subscript $\perp $ from now on (unless specified otherwise) following
the existing literature conventions. As in the main text, we need to switch
now from the space-time of Minkowski signature to that of Euclidean
signature. As it is typically done in quantum mechanics and quantum field
theories, this is achieved by replacing $t$ by $-i\tau $ resulting in
replacement of the factor $iS$ \ in the exponent of the path integral by $%
-S. $ The Euclideanized action acquires the following form 
\begin{equation}
S[\mathbf{A}]=\frac{1}{2}\tint\limits_{-\infty }^{\infty
}dt\tint\limits_{Y}dv[\mathbf{\dot{A}}^{2}+\left( \mathbf{\nabla }\times 
\mathbf{A}\right) ^{2}]  \tag{A.5}
\end{equation}%
in which $\frac{1}{2}\tint\limits_{Y}dv[\mathbf{\dot{A}}^{2}+\left( \mathbf{%
\nabla }\times \mathbf{A}\right) ^{2}]=\frac{1}{2}\tint\limits_{Y}dv[\mathbf{%
E}^{2}+\mathbf{B}^{2}]$ is easily recognizable as eq.(2.11) of the main text.

Use of eq.(A.3) allows us to discuss the electric-magnetic duality. Deser
and Teitelboim (1976) noticed that although Maxwell's equations without
sources are invariant with respect to the duality transformation: $\mathbf{%
E\rightarrow B,B\rightarrow -E,}$ the action functional (A.1) is not. \
Naively, this fact is not causing any problems when Maxwell's equations are
recovered from the \ action functional $S$ variationally. Nevertheless they
and subsequent researchers indicated that such non invariance of $S$ is
highly undesirable. To fix the problem we follow the paper by Pakman (Pakman
2000). Later on Bunster and Henneaux (Bunster and Henneaux 2011) had reached
the same conclusions.

For Maxwellian fields the action given in eq.(2.5) can be written as 
\begin{equation}
S[\mathbf{F}]=-tr\tint\limits_{M}(\mathbf{F}\wedge \ast \mathbf{F})=-\frac{1%
}{4}tr\dint\limits_{M}dvF^{\mu \upsilon }F_{\mu \upsilon }  \tag{A.6}
\end{equation}%
with $F_{\mu \upsilon }[\mathbf{A}]=\partial _{\mu }A_{\upsilon }-\partial
_{\upsilon }A_{\mu }$. From it we get the equations of motion 
\begin{equation}
\partial _{\mu }F^{\mu \upsilon }[A]=0  \tag{A.7}
\end{equation}%
and the Bianchi identities 
\begin{equation}
\partial _{\mu }\tilde{F}^{\mu \upsilon }[A]=0  \tag{A.8}
\end{equation}%
with $\tilde{F}^{\mu \upsilon }[A]$ being dual of $F_{\mu \upsilon }[\mathbf{%
A}].$ That is using eq.(2.13) we can rewrite the action $S$ as well as 
\begin{equation*}
S[\mathbf{F}]=-\frac{1}{4}tr\dint\limits_{M}dv\tilde{F}^{\mu \upsilon }%
\tilde{F}_{\mu \upsilon }.
\end{equation*}%
Next, we introduce new variables $Z$ such that $F_{\mu \upsilon
}[Z]=\partial _{\mu }Z_{\upsilon }-\partial _{\upsilon }Z_{\mu }.$
Evidently, it is possible the to interpret $\partial _{\mu }\tilde{F}^{\mu
\upsilon }[Z]=0$ as equations of motion while $\partial _{\mu }F^{\mu
\upsilon }[Z]=0$ as Bianchi identities. Using these results we can rewrite
the action $S$ in the form 
\begin{equation}
S=-\frac{1}{8}\{S[A]+S[Z]\}  \tag{A.9}
\end{equation}%
But the Lagrangian of such defined action is easily recognizable as Ranada's
eq(5) of (Ranada 1989)! \ Furthermore, according to (Costa-Quintana and
Lopez -Aguillar 2012), in this action $\tilde{F}^{\mu \upsilon }[Z]$\ =$%
\frac{1}{2}\varepsilon ^{\mu \nu \rho \sigma }F_{\rho \nu }[A]$ , again, in
accord with\ Ranada (1989). Given this observation, we are not yet done.
Following Pakman we represent the action $S$ in still another form given by 
\begin{equation}
S[A_{\mu },F_{\mu \upsilon }]=\dint\limits_{M}dv[\frac{1}{4}F^{\mu \upsilon
}F_{\mu \upsilon }-\frac{1}{2}F_{\mu \upsilon }(\partial _{\mu }A_{\upsilon
}-\partial _{\upsilon }A_{\mu })]  \tag{A.10}
\end{equation}%
where $A_{\mu }$ and $F_{\mu \upsilon }$ are being treated as independent
variables. In view of eq.(A.3) it is convenient to introduce $\mathbf{E}%
_{\perp }=\nabla \times \mathbf{Z.}$ It is also convenient now to rewite
eq.(A.4) as%
\begin{eqnarray}
S[\mathbf{A}_{\perp }] &=&\frac{1}{2}\tint\limits_{-\infty }^{\infty
}dt\tint\limits_{Y}dv[-\mathbf{E}_{\perp }\cdot \mathbf{\dot{A}}_{\perp
}-\left( \mathbf{\nabla }\times \mathbf{A}_{\perp }\right) ^{2}]  \notag \\
&=&-\frac{1}{2}\tint\limits_{-\infty }^{\infty }dt\tint\limits_{Y}dv[(\nabla
\times \mathbf{Z)\cdot \dot{A}}_{\perp }+\left( \mathbf{\nabla }\times 
\mathbf{A}_{\perp }\right) ^{2}]  \TCItag{A.11}
\end{eqnarray}%
But, in view of eq.(A.9) we can rewrite this result also as 
\begin{equation}
S[\mathbf{A}_{\perp },\mathbf{Z}]=-\frac{1}{2}\tint\limits_{-\infty
}^{\infty }dt\tint\limits_{Y}dv[(\nabla \times \mathbf{Z)\cdot \dot{A}}%
_{\perp }+\left( \mathbf{\nabla }\times \mathbf{A}_{\perp }\right)
^{2}-(\nabla \times \mathbf{A}_{\perp }\mathbf{)\cdot \dot{Z}+}\left( 
\mathbf{\nabla }\times \mathbf{Z}\right) ^{2}]  \tag{A.12}
\end{equation}%
To check correctness of this result we can perform independent variation
with respect to \textbf{Z} and \textbf{A}$_{\perp }.$ Using integration by
parts several times we arrive at 
\begin{equation*}
\mathbf{\dot{E}}=\nabla \times \mathbf{\dot{Z}=\nabla \times \nabla \times A}%
_{\perp }\mathbf{=\nabla \times B;\dot{B}=}\nabla \times \mathbf{\dot{A}}%
_{\perp }\mathbf{=-\nabla \times \nabla \times Z=-\nabla \times E.}
\end{equation*}%
These are Maxwell's equations as required. The action functional, eq.(A.12),
is manifestly invariant with respect to duality transformations. Finally,
using these results and eq.s(12-13) of (Costa-Quintana and Lopez -Aguillar
2012) we conclude that indeed the equations 
\begin{equation*}
\nabla \times \mathbf{\dot{Z}=\nabla \times \nabla \times A}_{\perp }\text{
and }\nabla \times \mathbf{\dot{A}}_{\perp }\mathbf{=-\nabla \times \nabla
\times Z}
\end{equation*}%
are describing electromagnetic fields originating from a dyon. More on dyons
can be found, for example, in (Negi and Dehnen 2011).

\bigskip

\textit{Appendix B 3+1 decomposition of Y-M fields\bigskip }

In this appendix we would like to provide details of
derivation/justification of eq.(2.32). For this purpose, following Donaldson
(2002) we need first to describe the 4-manifolds to be used. These are
different from \textbf{R}$^{4}$ (or $S^{4})$ used in physics literature
(Polyakov 1987). Accordingly, the fiber bundle in the present case is also
different from that used in physics literature.

Let Y$_{i}$ be a collection of compact Riemannian 3-manifolds\footnote{%
That is 3-manifolds with Euclidean-type signature}. The index $i(i=1,...,n)$
may contain just 1 entry. Let $U_{i}=Y_{i}\times (0,\infty ).$ It is to be
called "half-tube". To connect such constructed half tube with the design
known in physics, it is helpful to notice that when $Y_{i}=S^{3}$ the half
tube $S^{3}\times (0,\infty )$ is conformally equivalent to the punctured
4-ball $\mathbf{B}^{4}\smallsetminus \{0\},$ that is to \textbf{R}$^{4}.$
The $S^{3}$ is called the "cross-section". More \ generally, it is possible
to construct 4-manifold $X$ with tubular ends \ such that each of these ends
is having $S^{3}$ as a cross-section. Then, $X$ is conformally equivalent \
to a punctured manifold $\tilde{X}\smallsetminus \{p_{1},...,p_{n}\}$ where $%
\tilde{X}$ is being compact (previously we had $\mathbf{B}^{4}\smallsetminus
\{0\}$ and $\mathbf{B}^{4}$ respectively). As is well known, both the norm,
eq.(2.5), and the instanton, eq.(2.7), are conformally invariant (Donaldson
and Kronheimer 1990)

The purpose of such constructed 4-manifolds is exactly the same as in 3+1
decomposition of spacetimes \ used in general relativity\footnote{%
For a quick introduction to this topic our readers can consult (Kholodenko
2011), version 1.}. \ Specifically, it is possible to pass from connections
on the tube $Y\times \mathbf{R}$ to one -parameter family of connections on $%
Y$. That is \textsl{the} \textsl{Euclidean time parametrizes connections on} 
$Y.$ Thus, locally, a connection \textbf{A} over tube is given by the
connection matrix 
\begin{equation}
\mathbf{A}=A_{0}dt+\tsum\limits_{i=1}^{3}A_{i}dy^{i}.  \tag{B.1}
\end{equation}%
In this expression both $A_{0}$ and $A_{i}$ are functions of $%
t,y_{1},y_{2},y_{3.}.$ In the temporal gauge $A_{0}=0$ the Euclidean time $t$
becomes a parameter for $A_{i}^{\prime }s.$ The Hodge operator $\ast $
initially defined on 4-manifold can now be adopted to 3-manifold $Y$ via the
following prescription. Let $\phi $ be 1-form on $Y$. Construct next a
2-form $dt\wedge \phi $ on $M$ and consider its dual $\ast (dt\wedge \phi ),$
then since $\ast $ is acting on half-tube, it is possible to write%
\begin{equation}
\ast (dt\wedge \phi )=\ast _{3}\phi  \tag{B.2}
\end{equation}%
where $\ast _{3}$ is the Hodge operator acting in $Y$ only. \ Based on this
result, all anti-self-dual forms can be brought into form%
\begin{equation}
\Phi =\phi \wedge dt+\ast _{3}\phi .  \tag{B.3}
\end{equation}%
To check if this is the desired form, we have to demonstrate that $\ast \Phi
=-\Phi .$ But $\ast \Phi =\ast \lbrack \phi \wedge dt+\ast _{3}\phi ]=\ast
(\phi \wedge dt)+\ast \ast (dt\wedge \phi )=-\ast (dt\wedge \phi )-\phi
\wedge dt=-\ast _{3}\phi -\phi \wedge dt$ QED.

\ Now we take the 1-form given in eq.(B.1) and apply $d$ operator to it.
Thus, we obtain,%
\begin{eqnarray}
\mathbf{F} &=&\tsum\limits_{i=1}^{3}\left( \frac{\partial }{\partial t}%
A_{i}\right) dt\wedge dy^{i}+\tsum\limits_{i<j}(\frac{\partial }{\partial
y_{j}}A_{i}-\frac{\partial }{\partial y_{i}}A_{j})dy^{j}\wedge dy^{i}  \notag
\\
&=&\tsum\limits_{i=1}^{3}F_{0i}dt\wedge
dy^{i}+\tsum\limits_{i<j}F_{ij}dy^{j}\wedge dy^{i}  \notag \\
&=&\tsum\limits_{i=1}^{3}E_{i}dy^{i}\wedge dt-B_{1}dy^{2}\wedge
dy^{3}-B_{2}dy^{3}\wedge dy^{1}-B_{3}dy^{1}\wedge dy^{2},  \TCItag{B.4}
\end{eqnarray}%
in accord with eq.(2.1b). In arriving at this result we took into account
that $\mathbf{E}=-\frac{\partial }{\partial t}\mathbf{A}$ as discussed in
appendix A. Evidently, $\mathbf{F}=\Phi .$ But we \ just have demonstrated
that $\ast \Phi =-\Phi $. So, it remains to write down this result
explicitly. In Euclidean space the easiest way to obtain the desired result
is to consider it in the component-wise form. Specifically, we have to
consider only the following anti-self-dual conditions%
\begin{equation}
F_{01}=-F_{23},F_{02}=-F_{31},F_{03}=-F_{12}.  \tag{B.5}
\end{equation}%
By combining eq.s(B.4) and (B.5) we obtain 
\begin{equation}
-E_{i}=\frac{\partial }{\partial t}A_{i}=B_{i},i=1,2,3  \tag{B.6}
\end{equation}%
in accord with eq.(2.32). Finally, we take into account that: a) $\frac{%
\delta CS(\mathbf{A})}{\delta \mathbf{A}}=\mathbf{F}$ and b) $B_{i}=\ast
_{3}F_{0i},$ so that eq.(2.32) can now be rewritten as 
\begin{equation}
\frac{\partial }{\partial t}A_{i}=\ast _{3}F_{0i}  \tag{B.7}
\end{equation}%
This result is in accord with eq.(2.11) of Donaldson's book (Donaldson 2002)
where it was given without derivation.

The above results use essentially 3+1 decomposition of space-time. Use of
such a decomposition allows us to introduce the Hodge operator $\ast _{3}.$ 
\textsl{By construction}, \textsl{it will act the in} \textsl{the same way
both in spaces of Euclidean and Minkowski signature} as noticed already by
Donaldson\footnote{%
E.g. read page 35 (bottom) of (Donaldson 2002)}. The anti-self-duality eq.s
(B.6) and (B.7) reflect this fact. \ We would like to illustrate these
statements \ using \ Maxwell's equations written in 3 dimensional form.
These are given by\footnote{%
These two equations of the main text are reproduced here for reader's
convenience}%
\begin{equation}
\text{div}\mathbf{B}=0,\nabla \times \mathbf{E}=-\frac{\partial \mathbf{B}}{%
\partial t}\text{ (Bianchi identity }dF=0\text{),}  \tag{2.8a}
\end{equation}%
\begin{equation}
\text{div}\mathbf{E}=0,\nabla \times \mathbf{B}=\frac{\partial \mathbf{E}}{%
\partial t}\text{ (Equations of motion }d\ast F=0\text{).}  \tag{2.8b}
\end{equation}%
Let now $-\mathbf{E}=\mathbf{B}$ (see eq.(B.6)). Use this result in
eq.(2.8b) to obtain%
\begin{equation*}
\nabla \times \mathbf{B}+\frac{\partial \mathbf{B}}{\partial t}=0\text{ \
or, equivalently, }\nabla \times \mathbf{E}+\frac{\partial \mathbf{E}}{%
\partial t}=0.
\end{equation*}%
Since eq.(2.8a) can be rewritten as $\nabla \times \mathbf{E}+\frac{\partial 
\mathbf{\nabla \times A}}{\partial t}=0,$ this brings us back to the already
known result: $\mathbf{E}=-\frac{\partial }{\partial t}\mathbf{A,}$ \ thus
confirming correctness of the anti-self-duality requirement, eq.(B.6). Just
obtained results are outcomes of the correctly posed Cauchy problem for
Maxwellian fields.

\bigskip

\textit{Appendix C Some facts from hydrodynamics \ and magnetohydrodynamics
of \ ideal fluids}

\textit{\bigskip }

We shall be concerned only with the incompressible ideal fluids. The
incompressibility requires us to impose a constraint: $div\mathbf{v}=0$,
where \textbf{v} is fluid velocity. For simplicity, let the fluid density $%
\rho =1$ then, Euler's equation acquires the form%
\begin{equation}
\frac{\partial }{\partial t}\mathbf{v}+\mathbf{v}\cdot \nabla \mathbf{v}%
=-\nabla \mathcal{P}  \tag{C.1}
\end{equation}%
where $\mathcal{P}$ is the pressure. Since 
\begin{equation}
\frac{1}{2}\nabla \mathbf{v}^{2}=\mathbf{v\times \vec{\omega}+v}\cdot \nabla 
\mathbf{v}  \tag{C.2}
\end{equation}%
eq.(C.1) can be rewritten as 
\begin{equation}
\frac{\partial }{\partial t}\mathbf{v=v\times \vec{\omega}-\nabla (}\mathcal{%
P}\mathbf{+}\frac{\mathbf{v}^{2}}{2})  \tag{C.3a}
\end{equation}%
where the vorticity $\mathbf{\vec{\omega}=\nabla \times v.}$ This equation
can be interpreted electrodynamically if we notice that $\nabla \cdot 
\mathbf{\vec{\omega}=}\nabla \cdot (\nabla \mathbf{\times v)=}0\mathbf{.}$
Thus, we may formally identify $\mathbf{\vec{\omega}}$ with the magnetic
field $\mathbf{B.}$ Accordingly, the combination $-\mathbf{\nabla (}\mathcal{%
P}\mathbf{+}\frac{\mathbf{v}^{2}}{2})$ can be identified with the electric
field $\mathbf{E}$. Then, eq.(C.3a) acquires the form of Newton's equation \
for the particle of unit mass and charge moving under the influence of
Lorentz \ force, that is 
\begin{equation}
\frac{d}{dt}\mathbf{v}=\mathbf{E}+\mathbf{v\times B.}  \tag{C.4}
\end{equation}%
By applying the curl operator to eq.(C.3a) we obtain 
\begin{equation}
\frac{\partial }{\partial t}\mathbf{\vec{\omega}=\nabla \times (v\times \vec{%
\omega}).}  \tag{C.3b}
\end{equation}%
Since $\mathbf{\vec{\omega}\rightleftarrows B}$ we can rewrite eq.(C.3b) in
the form used in ideal magnetohydrodynamics (MHD)%
\begin{equation}
\frac{\partial }{\partial t}\mathbf{B=\nabla \times (v\times B).}  \tag{C.5}
\end{equation}%
Then, using \ Maxwell's equation $\nabla \times \mathbf{E}=-\dfrac{\partial 
\mathbf{B}}{\partial t}$ \ we can rewrite eq.(C.5) as%
\begin{equation}
\nabla \times (\mathbf{E+v\times B)=}0\mathbf{.}  \tag{C.6}
\end{equation}%
From here we obtain 
\begin{equation}
\mathbf{E+v\times B=-\nabla \Phi .}  \tag{C.7}
\end{equation}%
where $\Phi $ is some scalar potential. \ Multiplication of both sides by 
\textbf{B }produces 
\begin{equation}
\mathbf{B}\cdot \mathbf{\nabla \Phi =-E\cdot B}\text{.}  \tag{C.8}
\end{equation}%
Using this result we obtain as well (Boozer 2010) 
\begin{equation}
\mathbf{v=}\frac{\left( \mathbf{E}+\mathbf{\nabla \Phi }\right) \times 
\mathbf{B}}{\mathbf{B}^{2}}  \tag{C.9}
\end{equation}%
The null fields are obtained when $\mathbf{\nabla \Phi =0}$ (Irvine 2010).
Indeed, when this happens, \ eq.(C.8) produces $\mathbf{E}\cdot \mathbf{B}%
=0. $ It can be proven (Dubrovin et al 1984) that when $\mathbf{E}\cdot 
\mathbf{B}=0$ it is always possible to find a frame in which $\mathbf{E}^{2}=%
\mathbf{B}^{2}.$

\bigskip\ 

\textit{Appendix D} \ \textit{All about and around equation} $\mathbf{%
E+\nabla \Phi =0.}$ \bigskip

In Appendix A we decomposed vector fields into longitudinal and transverse
parts. That is we had $\mathbf{A}=\mathbf{A}_{\parallel }+\mathbf{A}_{\perp
} $ where $\mathbf{A}_{\parallel }=\nabla \varphi $ while $\mathbf{A}_{\perp
}=\nabla \times \mathbf{\tilde{A}.}$ It happens that there are other
decompositions as well. One of them is due to Clebsch (Lamb 1945). Such a
decomposition is especially useful when equations of hydrodynamics \ need to
be rewritten in terms of Hamiltonian equations of classical mechanics.
References (Seliger and Whitham 1968) and (Sudarshan and Mukunda 1974)
contain excellent treatments of such Hamiltonization protocol. Clebsch
variables were also used in Ranada's paper (Ranada 1989)\footnote{%
Albeit for a different reason} as well as in most of recent papers on
optical knots. Here we discuss them in the context of equation $\mathbf{%
E+\nabla \Phi =0.}$

\ The idea of Clebsch lies in the following. Associate with the vector field 
$\mathbf{A(r)}$\textbf{\ }three scalar\textbf{\ }functions $\varphi (\mathbf{%
r}),\alpha (\mathbf{r})$ and $\beta (\mathbf{r})$ so that $\mathbf{A}=\nabla
\varphi +\alpha \nabla \beta .$ Then, $\nabla \times \mathbf{A=\nabla }%
\alpha \mathbf{\times \nabla }\beta $. It is permissible to make these
scalar functions dependent on time $t$. \ This is done with the purpose of
relating the velocity \textbf{v} in eq.(C.3b)and (C.5) \ with Clebsch
scalars. \ Suppose that the vortex tube (Section 4) is described in terms of
an equation $f(t,x,y,z)=0$, then it is possible to define the normal $%
\mathbf{N}$ at each point of this tube: $\mathbf{N}=$\textbf{e}$_{x}\frac{df%
}{dx}$ +\textbf{e}$_{y}\frac{df}{dy}+$\textbf{e}$_{z}\frac{df}{dz}.$ The
vortex tube can then be described in terms of the equation (Lamb 1945,
Saffman 1995) $\mathbf{\vec{\omega}\cdot N=0}$ or, 
\begin{equation}
\lbrack \omega _{x}\frac{\partial }{\partial x}+\omega _{y}\frac{\partial }{%
\partial y}+\omega _{z}\frac{\partial }{\partial z}]f=0  \tag{D.1a}
\end{equation}%
which is typically \ can be re written as 
\begin{equation}
\frac{dx}{\omega _{x}}=\frac{dy}{\omega _{y}}=\frac{dz}{\omega _{z}}=dt\text{
or \ as }\frac{d\mathbf{r}}{dt}=\mathbf{\vec{\omega}(r(}t\mathbf{)).} 
\tag{D.1b}
\end{equation}%
But, $\mathbf{\vec{\omega}=\nabla \times v}$ and also, $\mathbf{\vec{\omega}%
(r(}t\mathbf{))=\nabla }\alpha (\mathbf{r}(t))\mathbf{\times \nabla }\beta (%
\mathbf{r}(t)).$ Therefore,%
\begin{equation}
\frac{d}{dt}\alpha (\mathbf{r}(t))=\mathbf{\nabla }\alpha (\mathbf{r}%
(t))\cdot \frac{d\mathbf{r}}{dt}=\mathbf{\nabla }\alpha (\mathbf{r}(t))\cdot 
\mathbf{\vec{\omega}(r(}t\mathbf{))=}0  \tag{D.2a}
\end{equation}%
and 
\begin{equation}
\frac{d}{dt}\beta (\mathbf{r}(t))=\mathbf{\nabla }\beta (\mathbf{r}(t))\cdot 
\frac{d\mathbf{r}}{dt}=\mathbf{\nabla }\beta (\mathbf{r}(t))\cdot \mathbf{%
\vec{\omega}(r(}t\mathbf{))=}0  \tag{D.2b}
\end{equation}%
These results are compatible with those discussed in section 4.1. Indeed,
since $\dfrac{d\mathbf{r}}{dt}=\mathbf{v}$ and, in view of eq.(D.1b), we
obtain: $\mathbf{v}=\nabla \times \mathbf{v.}$ This is Beltrami-type
equation. \ The obtained results suggest that both Clebsch scalars $\alpha $
and $\beta $ are constants of motion. The vortex orbit $\mathbf{r}(t)$ is
contained in the level sets of both $\alpha (\mathbf{r}(t),t)$ and $\beta (%
\mathbf{r}(t),t).$ This can be alternatively stated as the requirement that 
\textbf{v} is determined by both equations 
\begin{equation}
\frac{\partial }{\partial t}\alpha +\mathbf{v}\cdot \nabla \alpha =0, 
\tag{D.3a}
\end{equation}%
and 
\begin{equation}
\frac{\partial }{\partial t}\beta +\mathbf{v}\cdot \nabla \beta =0. 
\tag{D.3b}
\end{equation}
Since in the MHD laguage $\mathbf{B}=\nabla \alpha \times \nabla \beta ,$ by
multiplying eq.(D.3a) by $\nabla \beta $ and eq.(D.3b) by $\nabla \alpha $
and by subtracting the 2nd equation from the 1st, the following result is
obtained (Stern 1970)%
\begin{equation}
\mathbf{v}\times \mathbf{B=}\nabla \beta \left( \frac{\partial }{\partial t}%
\alpha \right) -\nabla \alpha \left( \frac{\partial }{\partial t}\beta
\right) .  \tag{D.4}
\end{equation}%
Now, we obtain as well%
\begin{eqnarray*}
\frac{\partial }{\partial t}\mathbf{B} &\mathbf{=}&\nabla \left( \frac{%
\partial }{\partial t}\alpha \right) \times \nabla \beta +\nabla \alpha
\times \nabla \left( \frac{\partial }{\partial t}\beta \right) \\
&=&\nabla \times \lbrack \nabla \beta \left( \frac{\partial }{\partial t}%
\alpha \right) -\nabla \alpha \left( \frac{\partial }{\partial t}\beta
\right) ] \\
&=&\nabla \times (\mathbf{v}\times \mathbf{B)}
\end{eqnarray*}%
in accord with eq.(C.5). The map $(A_{x},A_{y},A_{z})$ $\rightarrow (\varphi
(\mathbf{r}),\alpha (\mathbf{r})$, $\beta (\mathbf{r}))$ is not bijective
(Yoshida 2009), (Sudarshan and Mukunda 1974) though. \ Indeed, \ if $\alpha
^{\prime }$\textbf{\ }and\textbf{\ }$\beta ^{\prime }$ are functions of $%
\alpha $ and $\beta $ such that the Jacobian $\frac{\partial (\alpha
^{\prime },\beta ^{\prime })}{\partial (\alpha ,\beta )}=1,$then $\alpha
^{\prime }$\textbf{\ }and\textbf{\ }$\beta ^{\prime }$ can be used instead
of $\alpha $\textbf{\ }and\textbf{\ }$\beta .$The requirement $\frac{%
\partial (\alpha ^{\prime },\beta ^{\prime })}{\partial (\alpha ,\beta )}=1$
is the same as used in canonical transformations of Hamiltonian mechanics
and it is, in fact a canonical transformation as further explained in the
main text. Furthermore, it is possible also to make time-depenent
transformations of $\alpha $\textbf{\ }and\textbf{\ }$\beta $ which leave 
\textbf{B} unchanged. These transformations replace more familiar (to
physics educated readers) canonical transformations \ by the less familiar
contactomorphic transformations (Kholodenko 2013, Geiges 2008) discussed in
the main text. Boozer (2010) noticed that eq.(C.7) is well behaved only if $%
\mathbf{B}\cdot \nabla \Phi =-\mathbf{E}\cdot \mathbf{B.}$ In such a case
eq.(C.9) is obtained. Only if $\nabla \Phi =0$ we reobtain back results used
in (van Enk 2013) and (Irvine 2010) leading to null fields.

Because Clebsch transformations belong to a special case of Galilei-type
transformation as explained by Sudarshan and Mukunda (1974), the choice $%
\nabla \Phi =0$ corresponds to a choice of a special reference frame. \
Another choice of frame in which the vortex is not moving leads to $\mathbf{%
E+\nabla \Phi =}0$ as required for the validity of arguments of section 4.2.
\ Evangelidis (1988) demonstrated that: a) the force-free condition
eq.(4.5b) is both gauge and relativistically invariant; b) if
electromagnetic field is described by the pre assigned pair (\textbf{E}, 
\textbf{B}) in a given reference frame, it is always possible to find
another reference frame in which the electric field \textbf{E} is parallel
to the magnetic field \textbf{B. }Such a frame should move with velocity 
\textbf{v} (with respect to the frame with pre-assigned pair (\textbf{E}, 
\textbf{B})) determined by 
\begin{equation}
\mathbf{v}=(\mathbf{E\times B)}\frac{c}{2}\frac{\left\vert \mathbf{E}%
^{2}\right\vert +\left\vert \mathbf{B}^{2}\right\vert }{\sqrt{\left\vert 
\mathbf{E}^{2}\right\vert \left\vert \mathbf{B}^{2}\right\vert -\left( 
\mathbf{E}\cdot \mathbf{B}\right) ^{2}}}\{1\pm \frac{\sqrt{\left( \left\vert 
\mathbf{E}^{2}\right\vert -\left\vert \mathbf{B}^{2}\right\vert \right)
^{2}+4\left( \mathbf{E}\cdot \mathbf{B}\right) ^{2}}}{\left\vert \mathbf{E}%
^{2}\right\vert +\left\vert \mathbf{B}^{2}\right\vert }\}  \tag{D.6}
\end{equation}%
In the case if the preassigned pair (\textbf{E}, \textbf{B}) is such that $%
\left\vert \mathbf{E}^{2}\right\vert -\left\vert \mathbf{B}^{2}\right\vert
=0,$ and $\mathbf{E}\cdot \mathbf{B=0}$ we obtain: $\mathbf{v}=c$. Here $c$
is the speed of light. This is the case of null fields discussed in the
text. \ This conclusion is in accord with that presented in the paper by van
Enk (2013). The frame in which \textbf{E} is parallel to \textbf{B} was
suggested by Chu and Ohkawa (1982) who demonstrated that in this frame the
force-free eq.(4.5) holds true. Gray (1992) suggested physical conditions
under which \ one can create a situation in which \textbf{E} is parallel to 
\textbf{B. }He also suggested (in a way different from that discussed in
Section 4) that this condition leads to the localized configurations of the
electomagnetic field. Evangelidis (1988) obtained his results independently
from Brownstein (1987) who also came to the same conclusions using different
arguments.

\bigskip

\bigskip

\bigskip {\huge \ }

\bigskip \textbf{References\bigskip }

Arnol'd V 1989 \textit{Mathematical Methods of Classical Mechanics}

\ \ \ \ \ \ \ \ \ \ \ \ \ \ (Berlin: Springer-Verlag)

Arnol'd V and Khesin B \ 1998 \textit{Topological Methods in Hydrodynamics}

\ \ \ \ \ \ \ \ \ \ \ \ \ \ (Berlin: Springer-Verlag)

Arnol'd V \ 1984 \textit{Catastrophe Theory} (Berlin: Springer-Verlag)

Arnol'd V 1986 \ \ First steps in symplectic topology,

\ \ \ \ \ \ \ \ \ \ \ \ \ \ \ \ \textit{Russian Mathematical Surveys} 
\textbf{41} 1-21

Bialynicki-Birula I and Bialynicki-Birula Z 2013 The role of the
Riemann-Silberstein

\ \ \ \ \ \ \ \ \ \ \ \ \ \ \ \ vector in classical and quantum theories of
electromagnetism

\ \ \ \ \ \ \ \ \ \ \ \ \ \ \ \ \textit{J.Phys. A} \textbf{46} 053001

Birman J and Williams R 1983 Knotted periodic orbits in dynamical systems

\ \ \ \ \ \ \ \ \ \ \ \ \ \ \ \ \textit{Topology} \textbf{22} 47-82\ \ \ \ \
\ \ \ \ \ \ \ \ 

Boozer A \ 2010 \ Mathematics and Maxwell's equations

\ \ \ \ \ \ \ \ \ \ \ \ \ \ \ \ \ \textit{Plasma Phys.Control Fusion} 
\textbf{52} 124002

Bott R and Tu L 1982 \textit{Differential Forms in Algebraic Topology}

\ \ \ \ \ \ \ \ \ \ \ \ \ \ \ \ \ (Berlin: Springer-Verlag)

Bretheron F \ 1970 A note on Hamilton's principle for perfect fluids

\ \ \ \ \ \ \ \ \ \ \ \ \ \ \ \ \ \ \textit{J.Fluid Mech.}\textbf{44} 19-31

Brownstein K 1987 Transformation properties of the equation $\nabla \times 
\mathbf{V}=k\mathbf{V}$

\ \ \ \ \ \ \ \ \ \ \ \ \ \ \ \ \ \ \textit{Phys.Rev.} A \textbf{35}
4856-4858

Bunster C and Henneaux M 2011 Can (electric-magnetic) \ duality be gauged?

\ \ \ \ \ \ \ \ \ \ \ \ \ \ \ \ \ \ \ \textit{Physical Review D} \textbf{83}
045031\ \ 

Costa-Quintana J and Lopez-Aguilar F 2012 Extended Lagrangian formalisms for

\ \ \ \ \ \ \ \ \ \ \ \ \ \ \ \ \ \ \ dyons and some applications to solid
systems under external fields

\ \ \ \ \ \ \ \ \ \ \ \ \ \ \ \ \ \ \ \ \textit{Ann.Phys.} \textbf{327}
1948-1961

Chu C and Ohkawa T 1982 Transverse electromagnetic waves with $\mathbf{E}%
\parallel \mathbf{B}$

\ \ \ \ \ \ \ \ \ \ \ \ \ \ \ \ \ \ \ \ Phys.Rev.Lett.48 837-838

Chubukalo A, Espinosa A and Kosyakov B 2010 Self-dual electromagnetic fields

\ \ \ \ \ \ \ \ \ \ \ \ \ \ \ \ \ \ \ \ Am.J.Phys.\textbf{78} 858-861

Deser S and Teitelboim C 1976 Duality transformations of Abelian and

\ \ \ \ \ \ \ \ \ \ \ \ \ \ \ \ \ \ \ \ non-Abelian gauge fields \textit{%
Phys.Rev.D} \textbf{13} 1592-1597

Dennis M, King R, Jack B, O'Holleran K and Padgett M 2010 Isolated optical

\ \ \ \ \ \ \ \ \ \ \ \ \ \ \ \ \ \ \ \ \ vortex knots \textit{Nature Physics%
} \textbf{6} 118-121

Donaldson S 2002 \ \textit{Floer Homology Groups in Yang-Mills Theory}

\ \ \ \ \ \ \ \ \ \ \ \ \ \ \ \ \ \ \ \ \ (Cambridge UK: Cambridge
University Press)

Donaldson S and Kronheimer P 1990 \ \textit{The Geometry of Four-Manifolds}

\ \ \ \ \ \ \ \ \ \ \ \ \ \ \ \ \ \ \ \ \ (Oxford UK: Clarendon Press)

Dubrovin B, \ Fomenko A and Novikov S 1984\textit{\ Modern Geometry-Methods }

\ \ \ \ \ \ \ \ \ \ \ \ \ \ \ \ \ \ \ \ \ \textit{and Applications. Part I }%
(Berlin: Springer-Verlag)

Eguchi T , Gilkey P and Hanson \ A 1980 Gravitation, gauge theory and

\ \ \ \ \ \ \ \ \ \ \ \ \ \ \ \ \ \ \ \ \ differential geometry \textit{%
Phys.Rep}. \textbf{66} 213-293

Evangelidis E 1988 Electomagnetic fields satisfying the condition \textbf{B}$%
\wedge (\nabla \wedge \mathbf{B})=0$

\ \ \ \ \ \ \ \ \ \ \ \ \ \ \ \ \ \ \ \ \ \ \textit{Astrophysics and Space
Science} \textbf{143}, 113-121

Fischer A and Moncrief V 2001 The reduced Einstein equations and

\ \ \ \ \ \ \ \ \ \ \ \ \ \ \ \ \ \ \ \ \ \ \ conformal volume collaps of
3-manifolds\ 

\ \ \ \ \ \ \ \ \ \ \ \ \ \ \ \ \ \ \ \ \ \ \ \textit{Class.Quantum Grav}. 
\textbf{18} 4493-4515\ \ 

Finkelstein R 2006 Masses and Interactions of q-Fermionic Knots

\ \ \ \ \ \ \ \ \ \ \ \ \ \ \ \ \ \ \ \ \ \ \ \ \textit{Int.J.Mod.Phys.}A 
\textbf{21 }4269-4302

Floer A \ \ 1988 \ \ An instanton invariant for 3-manifolds \textit{%
Comm.Math.Phys}.

\ \ \ \ \ \ \ \ \ \ \ \ \ \ \ \ \ \ \ \ \ \ \ \ \ \ \textbf{118} 215-240

Frankel T 1997 \ \textit{The Geometry of Physics} \ (Cambridge UK: Cambridge
University Press)\ \ 

Geiges H 2008 \ \ \textit{An Introduction to Contact Topology}

\ \ \ \ \ \ \ \ \ \ \ \ \ \ \ \ \ \ \ \ \ \ \ \ \ \ (Cambridge UK: Cambridge
University Press)

Ghrist R, Holmes P and Sullivan M 1997 Knots and links in three dimensional
flows

\ \ \ \ \ \ \ \ \ \ \ \ \ \ \ \ \ \ \ \ \ \ \ \ \ \ \textit{Lecture Notes in
Mathematics }\textbf{1654}

\ \ \ \ \ \ \ \ \ \ \ \ \ \ \ \ \ \ \ \ \ \ \ \ \ \ (Berlin: Springer-Verlag)

Ghys E 2007 \ \ \ \ \ Knots and Dynamics

\ \ \ \ \ \ \ \ \ \ \ \ \ \ \textit{\ \ \ \ \ \ \ \ \ \ \ \ \ Proceedings of
the International Congress of Mathematicians}

\ \ \ \ \ \ \ \ \ \ \ \ \ \ \ \ \ \ \ \ \ \ \ \ \ \ \ \textit{Madrid Spain
2006 pp 247-277}

\ \ \ \ \ \ \ \ \ \ \ \ \ \ \ \ \ \ \ \ \ \ \ \ \ \ \ (Zurich: European
Mathematical Society)

Gilbert N and Porter T 1994 \textit{Knots and Surfaces}

\ \ \ \ \ \ \ \ \ \ \ \ \ \ \ \ \ \ \ \ \ \ \ \ \ \ \ \ (Oxford UK: Oxford
University Press)

Goncharov V and Pavlov V 1997 Some remarks on the physical foundation

\ \ \ \ \ \ \ \ \ \ \ \ \ \ \ \ \ \ \ \ \ \ \ \ \ \ \ \ of the Hamiltonian
description of fluid motion

\ \ \ \ \ \ \ \ \ \ \ \ \ \ \ \ \ \ \ \ \ \ \ \ \ \ \ \ \textit{Eur.J.Mech.}
B \textbf{16} 509-512

Graham R and Heney F 2000 Clebsch representation near points where

\ \ \ \ \ \ \ \ \ \ \ \ \ \ \ \ \ \ \ \ \ \ \ \ \ \ \ \ the vorticity
vanishes \textit{Phys.Fluids} \textbf{12} 744-746

Gray J \ \ \ \ \ \ 1992 \ Electromagnetic waves with \textbf{E} parallel to 
\textbf{B}

\ \ \ \ \ \ \ \ \ \ \ \ \ \ \ \ \ \ \ \ \ \ \ \ \ \ \ \ \textit{J.Phys}.A 
\textbf{25} 5373-5376

Helmholtz H 1858 Uber Integrale, der hydrodynamischen Gleichungen,

\ \ \ \ \ \ \ \ \ \ \ \ \ \ \ \ \ \ \ \ \ \ \ \ \ \ \ \ \ welhe der
Wirbelbewegungen entsprehen

\ \ \ \ \ \ \ \ \ \ \ \ \ \ \ \ \ \ \ \ \ \ \ \ \ \ \ \ \ \textit{J.f\"{u}r
die Reine und Angewandte Mathematik} \textbf{55} 25-55

Huang K 1982 \ \textit{Quarks,Leptons and Gauge Fields}

\ \ \ \ \ \ \ \ \ \ \ \ \ \ \ \ \ \ \ \ \ \ \ \ \ \ \ \ (Singapore:World
Scientific)

Irvine W 2010 \ Linked and knotted beams of light, conservation of helicity

\ \ \ \ \ \ \ \ \ \ \ \ \ \ \ \ \ \ \ \ \ \ \ \ \ \ \ \ and the flow of null
electromagnetic fields

\ \ \ \ \ \ \ \ \ \ \ \ \ \ \ \ \ \ \ \ \ \ \ \ \ \ \ \ \textit{J.Phys.A} 
\textbf{43} 385203

Jost J \ \ \ 2008 \ \ \ \ \ \textit{Riemannian Geometry and Geometric
Analysis}

\ \ \ \ \ \ \ \ \ \ \ \ \ \ \ \ \ \ \ \ \ \ \ \ \ \ \ (Berlin:
Springer-Verlag)

Kedia H, Bialynicki-Birula I, Peralta-Salas D and Irvine W 2013

\ \ \ \ \ \ \ \ \ \ \ \ \ \ \ \ \ \ \ \ \ \ \ \ \ \ \ \ Tying knots in light
fields \textit{Phys.Rev.Lett} \textbf{111} 150404

Kholodenko A \ 2013 \textit{Applications of Contact Geometry and Topology in
Physics}

\ \ \ \ \ \ \ \ \ \ \ \ \ \ \ \ \ \ \ \ \ \ \ \ \ \ \ \ \ (Singapore:World
Scientific)

Kholodenko A 2011 Gravity assisted solution of the mass gap problem for pure

\ \ \ \ \ \ \ \ \ \ \ \ \ \ \ \ \ \ \ \ \ \ \ \ \ \ \ \ Yang-Mills fields \ 
\textit{International Journal of Geometric Methods}

\ \ \ \ \ \ \ \ \ \ \ \ \ \ \ \ \ \ \ \ \ \ \ \ \ \ \ \textit{\ in Modern
Physics }\textbf{8} 1355-1418

Kholodenko A 2008 Towards physically motivated proofs of the Poincar$%
e^{^{\prime }}$ and

\ \ \ \ \ \ \ \ \ \ \ \ \ \ \ \ \ \ \ \ \ \ \ \ \ \ \ \ \ \ geometrization
conjectures \textit{J.Geom.Phys}. \textbf{58} 259-290

Kleckner D and Irvine W 2013\ Creation and dynamics of knotted vortices in
fluid\ \ 

\ \ \ \ \ \ \ \ \ \ \ \ \ \ \ \ \ \ \ \ \ \ \ \ \ \ \ \ \ \ \ \textit{Nature
Physics} \textbf{9} 253-258

Kuznetsov E and Mikhailov A 1980 On the topological meaning of canonical

\ \ \ \ \ \ \ \ \ \ \ \ \ \ \ \ \ \ \ \ \ \ \ \ \ \ \ \ \ \ Clebsch
variables \textit{Phys.Lett}. A \textbf{77} 37-38

Lamb \ H \ \ \ 1945 \ \ \textit{Hydrodynamics }(New York: Dower Publications)

Landau L and Lifshitz E 1975 \textit{Classical Theory of Fields}

\ \ \ \ \ \ \ \ \ \ \ \ \ \ \ \ \ \ \ \ \ \ \ \ \ \ \ \ \ (Oxford UK: Reed
Educational and Professional Publishing Ltd.)

Lomonaco S 1996 The modern legacies of Thompson's atomic vortex theory in

\ \ \ \ \ \ \ \ \ \ \ \ \ \ \ \ \ \ \ \ \ \ \ \ \ \ \ \ \ classical
electrodynamics in \textit{The Intrface of Knots and Physics}

\ \ \ \ \ \ \ \ \ \ \ \ \ \ \ \ \ \ \ \ \ \ \ \ \ \ \ \ \textit{\ pp 145-156 
}(Providence RI, AMS Publishers)

Manton N and Sutcliffe P 2007 \textit{Topological Solitons}

\ \ \ \ \ \ \ \ \ \ \ \ \ \ \ \ \ \ \ \ \ \ \ \ \ \ \ \ (Cambridge UK:
Cambridge University Press)

Marsden J and Weinstein A 1983 Coadjoint orbits, vortices, and Clebsch
variables for

\ \ \ \ \ \ \ \ \ \ \ \ \ \ \ \ \ \ \ \ \ \ \ \ \ \ \ \ incompressible
fluids \textit{Physica D} \textbf{7} 305-323

Mason L and Woodhouse N 1996 \textit{Integrability, Self-duality, and
Twistor Theory}

\ \ \ \ \ \ \ \ \ \ \ \ \ \ \ \ \ \ \ \ \ \ \ \ \ \ \ \ (Oxford UK:
Clarendon Press)

Mignaco J 2001 Electromagnetic duality, charges, monopoles, topology

\ \ \ \ \ \ \ \ \ \ \ \ \ \ \ \ \ \ \ \ \ \ \ \ \ \ \ \textit{Brazilian J.of
Physics} \textbf{31} 235-246

Nash C and Sen S 1983 \ \textit{Topology and Geometry for Physicists}

\ \ \ \ \ \ \ \ \ \ \ \ \ \ \ \ \ \ \ \ \ \ \ \ \ \ \ \ (New York: Academic
Press Inc.)

Negi O and Dehnen H 2011 Gauge formulation for two potential theory of dyons

\ \ \ \ \ \ \ \ \ \ \ \ \ \ \ \ \ \ \ \ \ \ \ \ \ \ \ \textit{%
Int.J.Theor.Phys}.\textbf{50} 2446-2459

Nego O and Dehnen H 2011 Gauge formulation for two potential theory of dyons

\ \ \ \ \ \ \ \ \ \ \ \ \ \ \ \ \ \ \ \ \ \ \ \ \ \ \ \textit{%
Int.J.Theor.Phys}. \textbf{50} 2446-2459

Newcomb W 1958 \ Motion of magnetic lines of force

\ \ \ \ \ \ \ \ \ \ \ \ \ \ \ \ \ \ \ \ \ \ \ \ \ \ \ \textit{Ann.Phys}.%
\textbf{3} 347-385

Nielsen H. and Olesen P 1973 Vortex-line models for dual strings

\ \ \ \ \ \ \ \ \ \ \ \ \ \ \ \ \ \ \ \ \ \ \ \ \ \textit{\ Nucl.Phys. B} 
\textbf{61} 45-61

Oh S-J \ \ \ 2013 \ \ \ Finite energy global well-posedness of the (3+1)
dimensional

\ \ \ \ \ \ \ \ \ \ \ \ \ \ \ \ \ \ \ \ \ \ \ \ \ \ \ Yang-Mills equations
using a novel Yang-Mills heat flow gauge

\ \ \ \ \ \ \ \ \ \ \ \ \ \ \ \ \ \ \ \ \ \ \ \ \ \ \ PhD Department of
Mathematics Princeton University

Packman R 2000 Schwarz-Sen duality made fully local

\ \ \ \ \ \ \ \ \ \ \ \ \ \ \ \ \ \ \ \ \ \ \ \ \ \ \ \ \ \textit{Phys.Lett.B%
} 474 309-314

Polyakov A 1987 \textit{Gauge Fields and Strings }

\ \ \ \ \ \ \ \ \ \ \ \ \ \ \ \ \ \ \ \ \ \ \ \ \ \ \ \ (New York: Harwood
Academic Publishers)

Ranada A 1989 \ \ A Topological theory of the electromagnetic field

\ \ \ \ \ \ \ \ \ \ \ \ \ \ \ \ \ \ \ \ \ \ \ \ \ \ \ \ \textit{%
Lett.Math.Phys.} \textbf{18} 97-106

Ranada A 1992 Topological electromagnetism

\ \ \ \ \ \ \ \ \ \ \ \ \ \ \ \ \ \ \ \ \ \ \ \ \ \ \ \textit{J.Phys.A} 
\textbf{25} 1621-1641

Ricca R 2001 \ \ \ \textit{An Introduction to the Geometry and Topology of
Fluid Flows}

\ \ \ \ \ \ \ \ \ \ \ \ \ \ \ \ \ \ \ \ \ \ \ \ \ \ \ (Boston: Kluver)

Ryder L 1980 \ \ \ \ Dirac monopoles and the Hopf map $S^{3}\rightarrow
S^{2} $

\ \ \ \ \ \ \ \ \ \ \ \ \ \ \ \ \ \ \ \ \ \ \ \ \ \ \ \textit{J.Phys.A} 
\textbf{13} 437-447

Saffman P 1995 \ \textit{Vortex Dynamics} (Cambridge UK: Cambridge
University Press)

Scott P \ \ \ \ \ 1983 \ The geometries of 3-manifolds

\ \ \ \ \ \ \ \ \ \ \ \ \ \ \ \ \ \ \ \ \ \ \textit{\ \ \ \ \ \ Bull. London
Math}.Soc. \textbf{15} 401-487

Seliger R and Whitham G 1968 Variational principles in continuum mechanics

\ \ \ \ \ \ \ \ \ \ \ \ \ \ \ \ \ \ \ \ \ \ \ \ \ \ \ \textit{Proc.Roy.Soc.A}
\textbf{305} 1-25

Stern D \ \ \ \ 1970 \ Euler Potentials

\ \ \ \ \ \ \ \ \ \ \ \ \ \ \ \ \ \ \ \ \ \ \ \ \ \ \ \textit{Am.J.Phys}. 
\textbf{38} 494-501

Streets J \ \ \ 2007 \ Ricci Yang-Mills flow

\ \ \ \ \ \ \ \ \ \ \ \ \ \ \ \ \ \ \ \ \ \ \ \ \ \ \ PhD Department of
Mathematics Duke University

Sudarshan E and Mukunda N 1974 \textit{Clssical Dynamics: A Moden Perspective%
}

\ \ \ \ \ \ \ \ \ \ \ \ \ \ \ \ \ \ \ \ \ \ \ \ \ \ \ (New York: John
wiley\&Sons Inc)

Thompson W 1868 On vortex motion \textit{Trans.Roy.Soc.Edinburg}\textbf{\ 25}
217-260

Thompson J 1883 \textit{A Treatise on the Motion of Vortex Rings }%
(London:Macmillan)

Thurston W 1976 Some simple examples of symplectic manifolds

\ \ \ \ \ \ \ \ \ \ \ \ \ \ \ \ \ \ \ \ \ \ \ \ \ \ \ \ \textit{%
Proc.Am.Math.Soc}. \textbf{55} 467-468

Trautman A 1977 Solutions of the Maxwell and yang-Mills equations associated

\ \ \ \ \ \ \ \ \ \ \ \ \ \ \ \ \ \ \ \ \ \ \ \ \ \ \ \ with Hopf fibrings 
\textit{International J.of Theoretical Physics} \textbf{16} 561-565

van Enk S 2013 The covariant description of electric and magnetic field lines

\ \ \ \ \ \ \ \ \ \ \ \ \ \ \ \ \ \ \ \ \ \ \ \ \ \ \ of null fields:
applications to Hopf-Ranada solutions

\ \ \ \ \ \ \ \ \ \ \ \ \ \ \ \ \ \ \ \ \ \ \ \ \ \ \ \textit{J.Phys.A} 
\textbf{46} 175204

Verjovsky A and Freyer R 1994 The Jones-Witten Invariant for flows on

\ \ \ \ \ \ \ \ \ \ \ \ \ \ \ \ \ \ \ \ \ \ \ \ \ \ \ a 3-dimensional
manifold

\ \ \ \ \ \ \ \ \ \ \ \ \ \ \ \ \ \ \ \ \ \ \ \ \ \ \ \textit{Comm.Math.Phys.%
} \textbf{163} 73-88

Witten E 1989 \ \ Quantum field theory and Jones polynomial

\ \ \ \ \ \ \ \ \ \ \ \ \ \ \ \textit{\ \ \ \ \ \ \ \ \ \ \ \ Comm.Math.Phys.%
} \textbf{121} 351-399

Yoshida Z 2009 \ Clebsch parametrization: Basic properties and remarks on

\ \ \ \ \ \ \ \ \ \ \ \ \ \ \ \ \ \ \ \ \ \ \ \ \ \ \ its applications 
\textit{J.Math.Phys.} \textbf{50} 113101

\ \ \ \ \ \ \ \ \ \ \ \ \ \ \ \ \ \ \ \ \ \ \ \ \ 

\ \ \ \ \ \ \ \ \ \ \ \ \ 

\ \ \ 

\ \ \ \ 

\ \ 

\end{document}